\documentclass[manuscript]{aastex}

\newcommand{\myemail}{antonios@star.ucl.ac.uk}

\shortauthors{Makrymallis et al.}

\usepackage{graphicx}
\usepackage{amssymb}
\usepackage{amsmath}
\usepackage[caption=false]{subfig}

\begin{document}

\title{Understanding the Formation and Evolution of Interstellar Ices: A Bayesian Approach }

\author{Antonios Makrymallis\altaffilmark{1}, Serena Viti}
\affil{Department of Physics \& Astronomy, University College London, WC1E 6BT, London UK }
\email{\myemail}
\altaffiltext{1}{Email Address: \myemail}

\begin{abstract}
Understanding the physical conditions of dark molecular clouds and star forming regions is an inverse problem subject to complicated chemistry 
that varies non-linearly with time and the physical environment.  In this paper we apply a Bayesian approach based on a Markov Chain Monte Carlo (MCMC)
method for solving the non-linear inverse problems encountered in astrochemical modelling. We use observations for ice and gas species in dark 
molecular clouds  and a time dependent, gas grain chemical model to infer the values of the physical and chemical parameters that characterize 
quiescent regions of molecular clouds. We show evidence that in high dimensional problems, MCMC algorithms provide a more efficient and complete 
solution than more classical strategies. The results of our MCMC method enable us to derive statistical estimates and uncertainties 
for the physical parameters of interest as a result of the Bayesian treatment.
\end{abstract}


\section{INTRODUCTION}

Molecular clouds are regions where extinction by dust is high ($A_V\gtrsim$5 mag), temperatures are low ($\sim$ 10 K), densities are high ($n_{\rm H}$ $\geq$ 10$^3$
cm$^{-3}$) and where most of the gas is molecular. They contain higher ($>$ 10$^4$ cm$^{-3}$) density structures \citep{Mye1983}, some of which may become gravitationally
unstable and initiate the early stages of star formation. Understanding the life cycle of dark molecular clouds is very important for 
comprehending star formation and for getting insight into the processes of the interstellar medium and to that extend galaxy formation. 
Molecules provide a paramount tool for the analysis of the chemical and physical conditions of star forming regions. Every stellar or planetary evolutionary stage 
is characterized by a chemical composition, which represents the physical processes of its phase.

In the dense cores of molecular clouds, molecules and atoms previously in the gas phase, deplete onto the dust grains. For each atom or molecule,
freeze out (or depletion) depends on a complicated time dependent, non linear chemistry that strongly depends on the physical
environment. It is difficult to quantify depletion observationally  \citep[e.g.][]{Chr2012}. CO emission can be used to infer the fraction of species
that is in the form of icy mantles, by taking the ratio of the observed CO to the expected abundance at a particular density in steady state, 
if freeze out did not occur \citep[e.g.][]{Cas1999}. This however, not only implies that the cores are in steady-state,
but also implies a knowledge of the H$_2$ density, as well as of the efficiency of the non thermal desorption mechanisms that can return
the depleted CO to the gas. Moreover, the CO depletion factor is not necessarily equivalent
to the molecular gas depletion factor, because different species freeze and desorb at different rates with different sticking coefficients, 
which are mostly unknown.

The detection of water ice mantles in cold dark interstellar clouds and star forming regions \citep{Obe2011b} provides us with direct evidence that surface
reactions on dust grains involving oxygen atoms make water molecules, which are then retained on the surface and make water ice.
Not all species undergo surface reactions when they stick to dust grains.
For example, CO sticks efficiently to surfaces at temperatures below $\sim$ 25 K and is found to be  abundant in the ices. Some of this CO 
can  be converted to other species.

The relatively high abundance of CO$_2$, CH$_3$OH, and H$_2$CO in ices \citep{Obe2011b, Whi2011}, relative to H$_2$O, in some clouds indeed suggest
that some processing of CO to these products is occurring, due possibly by irradiation, by cosmic rays or by photons generated by
cosmic rays inside the cloud. H$_2$CO and CH$_3$OH are stages in the surface hydrogenation of CO. Similarly, CO$_2$ can be
the result of oxygenation of CO:
\[
\rm{CO} + \rm{OH} \rightarrow \rm{CO}_2 + \rm{H}
\]

Some ices can be thermally returned to the gas phase when the gas temperature is higher than 20 K. At low gas temperatures
non-thermal desorption processes can also return molecules from solid to gas-phase (e.g. \citet{Rob2007}). However,
these mechanisms `compete' with those of freeze-out. The composition of the icy mantles is clearly a time-dependent process highly dependent on 
the initial conditions of the gas in any particular cloud. Hence, the ices on dust grain surfaces are of a mixed composition and may reflect the 
local conditions and evolutionary history. In some dark molecular clouds, the ices are abundant, indicating that non thermal 
desorption mechanisms may not be very efficient everywhere. The potential interconnection and linear or non-linear correlation of these parameters 
with each other or with extra unknown parameters augments our difficulty to determine and specify the parameter network. The large parameter space 
in combination with the number of parameters and the complexity of the physical system make the task of parameter estimation highly challenging.

The increasingly detailed observations of molecular clouds and star forming regions enable us to identify 
some of the most important processes at work. Chemical and radiative transfer models can transform molecular
observations into powerful diagnostics of the evolution and distribution of the molecular gas. The results of these models though, 
depend on a number of parameters or group of parameters that are most of the times poorly constrained. Moreover, deriving information about
molecular clouds using observational information and, even well established modeling codes, is an inverse problem that usually does
not fulfill Hadamard's \citep{Had1902} postulates of well-posedness. That is, it may not have a solution,
solutions might not be unique and/or might not depend continuously on the observational data.
The first and second postulates simply state
that for a well-posed problem a solution should exist and be unique. The third postulate holds when small changes in the observational data
result in small changes in the solution. As shown later in Section~\ref{sec-3.x1}, in typical astrochemical problems, only the first postulate holds 
and we usually have to deal with non linear ill-posed inverse problems.

Employing sampling algorithms is a  traditional approach to tackle inverse problems in many scientific fields with large parameter space.
Bayesian statistical techniques and Monte Carlo sampling methods such as  Markov Chain Monte Carlo (MCMC) algorithms
and Nested Sampling have flourished over the past decade in astrophysical data analysis \citep{Chr2000, For2005,Fit2007, Fer2008, Isa2009}.
A summary of a typical MCMC method and an application to quantify uncertainty in stellar parameters using stellar codes is given by \citet{Baz2012}. 
To our knowledge, MCMC methods have never been applied in the framework of parameter estimation through astrochemical modeling. In this paper we present 
a first astrochemical application of gas-grain chemical modeling, molecular abundances and a Bayesian statistical approach based on MCMC methodology.  

The motivation of the present paper is to solve the inverse problem of deriving the physical conditions in interstellar molecular clouds; 
in particular: the gas density, cosmic ray ionization rate, radiation field, the rate of collapse, the freeze-out rate and
non-thermal desorption efficiency. In Section~\ref{sec-2}, we formulate a typical inverse problem for interstellar molecular clouds and describe 
the bayesian method and the Metropolis-Hastings (MH) algorithm (an example of a wider class of MCMC techniques). In Section~\ref{sec-3}, we discuss the statistical results and 
the astrophysical consequences. Finally in Section~\ref{sec-4}, we present our conclusions.

\section{PARAMETER ESTIMATION}
\label{sec-2}
In this paper, we are interested in dense, cold, quiescent regions of molecular clouds where atoms and molecules in the gas phase freeze-out on to the dust grains.
The observed quantities are molecular abundances for solid and gas phase species. The parameters we want to estimate are the cloud density $n_H$, 
the cosmic ray ionization rate $\zeta$, radiation field rate $G_{\circ}$, the cloud collapse rate $C_f$ and three non thermal desorption efficiencies
$\epsilon$, $\phi$, $y$ presented in Section~\ref{sec-2.3}. 
Due to the nature of the addressed inverse problem, the theoretical and modeled relationship between the 
parameters and the observed data is highly non-linear. Therefore, we anticipate several degeneracies as well as a multi-modal and non-Gaussian
joint parameter distribution. Moreover, the parameters are not uniquely related to the observations. While the forward problem has (in deterministic physics)
a unique solution, the inverse problem does not. Different combinations of parameters can produce the same abundances. Furthermore, the possible combinations of
parameters are too many to permit an exhaustive search. 

Traditional approaches to tackle inverse problems of this nature fail to cope with these kind of issues. Methods based on searching iteratively to minimize an 
appropriate distance such as the $\rm{\chi}^2$ error, can be stuck in local minimum and give degenerate solutions. Alternative approaches to aim for a 
global solution such as simulated annealing would have some benefits, but since we are not just looking for the global optimum 
of our target distribution, the most comprehensive view is obtained by a Bayesian Monte Carlo sampling method.
We selected the Bayesian MCMC approach against other methods that work equally well with complex and multimodal target distributions (e.g. Nested Sampling), since 
MCMC constitutes a benchmark algorithm in Monte Carlo sampling and parameter estimation problems. 

To overcome the challenges of an ill-posed nonlinear inverse problem we adopted a Bayesian approach based on the use of Metropolis-Hastings (MH) algorithm.
The Bayesian framework for inverse problems is based on systematic modeling of all errors and uncertainties from the Bayesian 
viewpoint. The potential of this approach to solve difficult inverse problems with high noise levels and serious model uncertainties
is much higher and also allows for prior information to be incorporated. The Bayesian solution is the whole posterior distribution of
the parameters and therefore, there is not only one solution, but a set of possible values. The advantage of MCMC approach is that there
is no restriction concerning the non-linearity of the model. Moreover, an appropriate tuning of the MCMC parameters allows the algorithm 
to explore all modes of the target distribution. Finally, even though it is still not feasible to do an exhaustive search through the parameter
space, MCMC methods can effectively explore the parameters’ joint posterior distribution, since model computations are concentrated around 
regions of interest in the parameters space.

\subsection{Bayesian Inverse Problem}

Our aim is to obtain information about physical parameters of a molecular cloud
$ \boldsymbol{\theta}=(\theta_1,\theta_2,...,\theta_k )$, while we measure molecular abundances 
$ \boldsymbol{\mathcal{Y}}=(\mathcal{Y}_1,\mathcal{Y}_2,...,\mathcal{Y}_n )$. These quantities are related to a (forward) function
$f(\cdot)$ which represents the physical and chemical processes in the cloud.  The main challenge is that there is no closed form function $f$ mapping the 
parameters to the observations, which could be inverted. However, given a set of parameters, estimated abundance values for the species of
interest can be computed with astrochemical models denoted here as $\mathcal{C}(\cdot)$. The addressed problem in our case is how to estimate
$\boldsymbol{\theta}$ from 
\begin{equation}
 \boldsymbol{\mathcal{Y}}=\mathcal{C}(\boldsymbol{\theta}) + \boldsymbol{\varepsilon}
\end{equation}
and according to \citet{Idi2008} this constitutes an inverse problem.
The error term $\boldsymbol{\varepsilon}$, represents both the observational noise and the modeling error between $\mathcal{C}(\cdot)$ and $f(\cdot)$.

We treat $\boldsymbol{\mathcal{Y}}$, $\boldsymbol{\theta}$ and $\boldsymbol{\varepsilon}$ as random variables and define the “solution” of the 
inverse problem to be the posterior probability distribution of the parameters given the observations. This allows to model the noise via its statistical properties,
even though we do not know the exact instance of the noise entering our data. We can also optionally specify a priori the form of solutions that we believe to be more likely,
through a prior distribution. Thereby, we can attach weights to multiple solutions which explain the data. This is the Bayesian approach to inverse problems.

Assume we have $K$ parameters $\theta_k$ and $N$ solid phase observable quantities $\mathcal{Y}_n$.
The error $\varepsilon_n$ on each observation $\mathcal{Y}_n$ is assumed to be normally distributed with variance $\sigma_n^2$. In addition, it is assumed that the 
observational errors are independent. The $\sigma_n^2$ is considered to correspond to the 
uncertainty on $\mathcal{Y}_n$, which is solely dictated by the observation. The probability density function of the errors is given by:
\begin{displaymath}
p_\varepsilon (\boldsymbol{\varepsilon}) = \prod_{n=1}^N \frac{1}{(2\pi)^\frac{1}{2}\sigma_n^2} \exp(\frac{\varepsilon_n^2}{2\sigma_n^2})
\end{displaymath}
Using (1), we can define the likelihood function $\mathbb{L}$ of observations given a model parametrized by a set of parameters as
\begin{displaymath}
 \mathbb{L}(\boldsymbol{\theta};\boldsymbol{\mathcal{Y}}) = p_\varepsilon(\boldsymbol{\mathcal{Y}}-\mathcal{C}(\boldsymbol{\theta}))= 
 \prod_{n=1}^N \frac{1}{(2\pi)^\frac{1}{2}\sigma_n^2} \times \exp(-\frac{1}{2}\sum_{n=1}^N [\frac{\mathcal{C}(\theta_n)-\mathcal{Y}_n}{2\sigma_n}] )
\end{displaymath}
In case any prior information about the unknown parameters is available, the Bayesian approach allows for this information to be 
taken into account. This information can be integrated through a prior probability distribution on the parameters, say $\pi(\boldsymbol{\theta})$. 
Then parameter estimation can be performed through the posterior probability distribution (PPD), using Bayes' rule
\begin{equation}
\pi(\boldsymbol{\theta}|\boldsymbol{\mathcal{Y}}) = \frac{\mathbb{L}(\boldsymbol{\theta};\boldsymbol{\mathcal{Y}})\pi(\boldsymbol{\theta)}}{m(\boldsymbol{\mathcal{Y}})}
\end{equation}
The PPD expresses our uncertainty about the parameters after considering the observations and any prior information. The denominator is simply a normalization factor.

In reality we are not able to access the whole posterior probability distribution. Therefore, computation of parameter estimates or 
uncertainties is a hard task.  MCMC methods are efficient methods that allow to sample from complex probability distributions
and approximate complex probability densities.

\subsection{Markov Chain Monte Carlo}
MCMC methods are a powerful class of algorithms that  produce random samples distributed according to the distribution of interest. 
The importance and efficiency of MCMC methods lies in the fact that these samples can be used to approximate the probability density of the distribution
by calculating it only for a feasible number of parameter values. Among the several implementations of possible algorithms, we employ a
MH sampling algorithm \citep{Gil1995}.  The MH algorithm will enable us to explore the parameter space and approximate
efficiently the PPD. A theoretical introduction on MCMC and MH is far beyond the scope of this paper. However, in Appendix A, we 
briefly describe the MH algorithm and how MCMC is employed for parameter estimation in our case.
Note that the tuning of the MH algorithm as described in Appendix A is very crucial when aiming to approximate possibly multi-modal and non-Gaussian
distribution, which is the case for this study.

\subsection{Parameter Space}
\label{sec-2.3}
The chemical modeling code used in this paper and denoted as $\mathcal{C}(\cdot)$ in (1) is the UCL\_CHEM time dependent gas-grain chemical code
\citep{Vit2004} and is briefly described in Appendix B and references therein. Note that for each set of parameters, 
$\mathcal{C}(\cdot)$ provide us with time series of chemical abundances. 
We choose to extract the chemical abundances of interest for the time points when the final density is reached and the cloud collapse has finished. Even though
we ignore the previous time points, the time dependancy is still taken into account and investigated through exploration of different final density values.

The parameters for our chemical modeling code create a nine dimensional parameter space (9D)
for molecular clouds as used in our MH and described in Table~\ref{tbl-1}:
\begin{displaymath}
\boldsymbol{\theta} = (n_H,\zeta,G_{\circ},C_f,fr,\epsilon, \phi, y, r),
\end{displaymath}
In a first attempt to employ a Bayesian approach for deriving branching ratios for poorly understood chemical 
reaction pathways, we also investigated the parameter $r$, which controls how much of the gas phase Oxygen turns into ice H$_2$O or ice OH.
Parameter $r$ reflects the percentage of O that turns into H$_2$O, so that $1-r$ reflects the percentage of Oxygen that turns into OH.
Desorption efficiencies resulting from H$_2$ formation on grains, direct cosmic ray heating and cosmic ray induced photodesorption are
determined by parameters $\epsilon$, $\phi$ and $y$, as introduced and studied by \citet{Rob2007}.  The freeze-out parameter in our code is effectively 
the sticking coefficient, a number in the range of $0-100\%$ that adjusts the rate per unit volume at which species deplete on the grain.
For the free-collapse to a particular $n_H$ we used the modified formula of \citet{Raw1992}, where parameter $C_f$ is considered to be a 
retardation factor with a value less than one, to roughly mimic the magnetic and/or rotational support, or an acceleration factor with a value greater 
than one to simulate a collapse faster than a free-fall (e.g. due to external pressure). 
Table~\ref{tbl-1} lists the set of physical parameters studied in this paper along with their definition domain $\mathbb{D}_{{\theta_k}}$. 
The joint definition domain $\mathbb{D}_{{\theta}}$ represents the parameter space to explore.
The selected domain limits refer to the theoretical range of possible values for molecular clouds where atoms and molecules deplete on to the dust,
ensuring though that extreme values are included. 


\subsection{Observational Constraints}

The observational constraints of our analysis are based on data from the existing literature. Even though in this application
we are primarily interested in ices, we include both gas phase and solid phase observations. To avoid confusion, we will denote with $\boldsymbol{\mathcal{Y}}$ 
a vector containing any observed quantity and if required we will specify whether we refer to solid phase or gas phase observations.

The solid phase observations include column densities and visual extinction data for molecular clouds in front of field stars. 
Such sources often provide suitable opportunities to observe and study ices in
quiescent regions of the clouds \citep[e.g.][]{Boo2011}. We used 31 observations of H$_2$O, CH$_3$OH, CO and CO$_2$
from 31 different regions of 16 different clouds found in literature and summarized by \citet{Whi2011}. 
The data suggest some abundance variation, which was attributed to different evolutionary stages for different clouds. 
The scope of this paper lie beyond studying the behavior of a specific cloud, but rather on how to get statistical insight into the dynamics 
of common cloud classes. Therefore, the observational data is transformed into fractional abundances with respect to total H nuclei and then 
the average value is computed and used for our analysis. 

In an attempt to minimize degeneracies we introduce additional gas phase abundances as an optional observational constraint. 
Due to the ill-posed nature of our problem, it is possible for our chemical model to end up with a solution space 
that fits perfectly the solid phase observations, but with gas phase abundances far from realistic. Hence, the addition of gas phase observations
can be considered as a mathematical regularization by introducing additional prior information. Prior information can be naturally integrated into our Bayesian
approach. The gas species observations were collected from more than one study, attempting to match the clouds, regions or evolutionary stage of the observational 
sources used for the solid phase species.
If we were to fit  observations of a particular source, then, ideally, every observational gas phase constraint should be able to contribute to the regularization 
of our methodology. However, as we are here only attempting at exploring a methodology, we found that three gas phase species were adequate to provide insight on 
the efficiency of gas phase species as a regularization factor.
Abundances for NH$_3$ and  N$_2$H$^+$ were collected from \citet{Joh2010}
while HCO$^+$ from \citet{Sch2002}. The gas phase observations are in the form of fractional abundances with 
respect to total hydrogen nuclei. 
Table~\ref{tbl-2} lists the average molecular abundances for all the species along with their uncertainties. We emphasize
again that the error on each of the observations $\mathcal{Y}_n$ is assumed to be normally distributed with a variance $\sigma_n^2$ that is determined solely
by the uncertainty reported in Table~\ref{tbl-2}. 

\subsection{Priors}

We run two identical sets of 8 MCMC chains that differ on the prior distribution information. For the first set, the prior information is non-informative and
 in the form of acceptable range of possible values. Therefore, $\pi(\boldsymbol{\theta})$ is just uniformly 
distributed on $\mathbb{D}_{{\theta}}$, as listed in Table~\ref{tbl-1}. Note that the observational data $\boldsymbol{\mathcal{Y}}$
refers only to the solid phase molecular abundances and in this case the gas phase species are ignored.
In the second case, the prior information includes the observational constraints from the gas phase species as well.
Let $\boldsymbol{\mathcal{Y}}$ now include all the observational constraints, $\boldsymbol{\mathcal{Y}}_s$ just the solid phase and $\boldsymbol{\mathcal{Y}}_g$ the gas phase observational constraints.
In that case, the PPD is defined as:
\begin{equation}
\pi(\boldsymbol{\theta}|\boldsymbol{\mathcal{Y}})= \pi(\boldsymbol{\theta}|\boldsymbol{\mathcal{Y}}_s,\boldsymbol{\mathcal{Y}}_g) = 
\frac{\pi(\boldsymbol{\mathcal{Y}}_s|\boldsymbol{\theta},\boldsymbol{\mathcal{Y}}_g)\pi(\boldsymbol{\theta}|\boldsymbol{\mathcal{Y}}_g)}{m(\boldsymbol{\mathcal{Y}})}
\end{equation}
 The prior information is simply the likelihood function $\mathbb{L}(\cdot)$ of  $\boldsymbol{\mathcal{Y}}_g$ given a model parametrized by $\boldsymbol{\theta}$, since:
\begin{align*}
\pi(\boldsymbol{\mathcal{Y}}_s|\boldsymbol{\theta},\boldsymbol{\mathcal{Y}}_g) &= \mathbb{L}(\boldsymbol{\theta};\boldsymbol{\mathcal{Y}}_s) \\
 \pi(\boldsymbol{\theta}|\boldsymbol{\mathcal{Y}}_g) &\propto \mathbb{L}(\boldsymbol{\theta};\boldsymbol{\mathcal{Y}}_g)\pi(\boldsymbol{\theta)}  
\end{align*}
Including prior information in this way is equivalent to attaching weight to the solutions that explain the gas phase as well as the solid phase chemistry. 

\subsection{Blind Benchmark Test}
In order to quantitatively investigate the effectiveness of our method to astrochemical problems we performed a benchmark test. This benchmark test is 
basically our Bayesian analysis applied this time on synthetic observations produced by UCL\_CHEM using a pre-defined set of parameters 
$\boldsymbol{\theta_T}$. Once we have our synthetic observations, we apply our methodology and analyse the 
results and whether the true parameters are recovered. Knowing the solution to this test a priori, allows us not only to validate the method,
but also to critically perceive the non linear and ill-posed nature of our problem. This discussion can be found in section 3.1. 
The reasoning behind the particular selection of parameters was a random choice not far from expected or well 
accepted values in the literature. The parameter values used in the test can be found in Table~\ref{tbl-x1}. 

\section{RESULTS}%
\label{sec-3}

\begin{figure}
\centering
\subfloat[]{
  \includegraphics[width=0.33\textwidth]{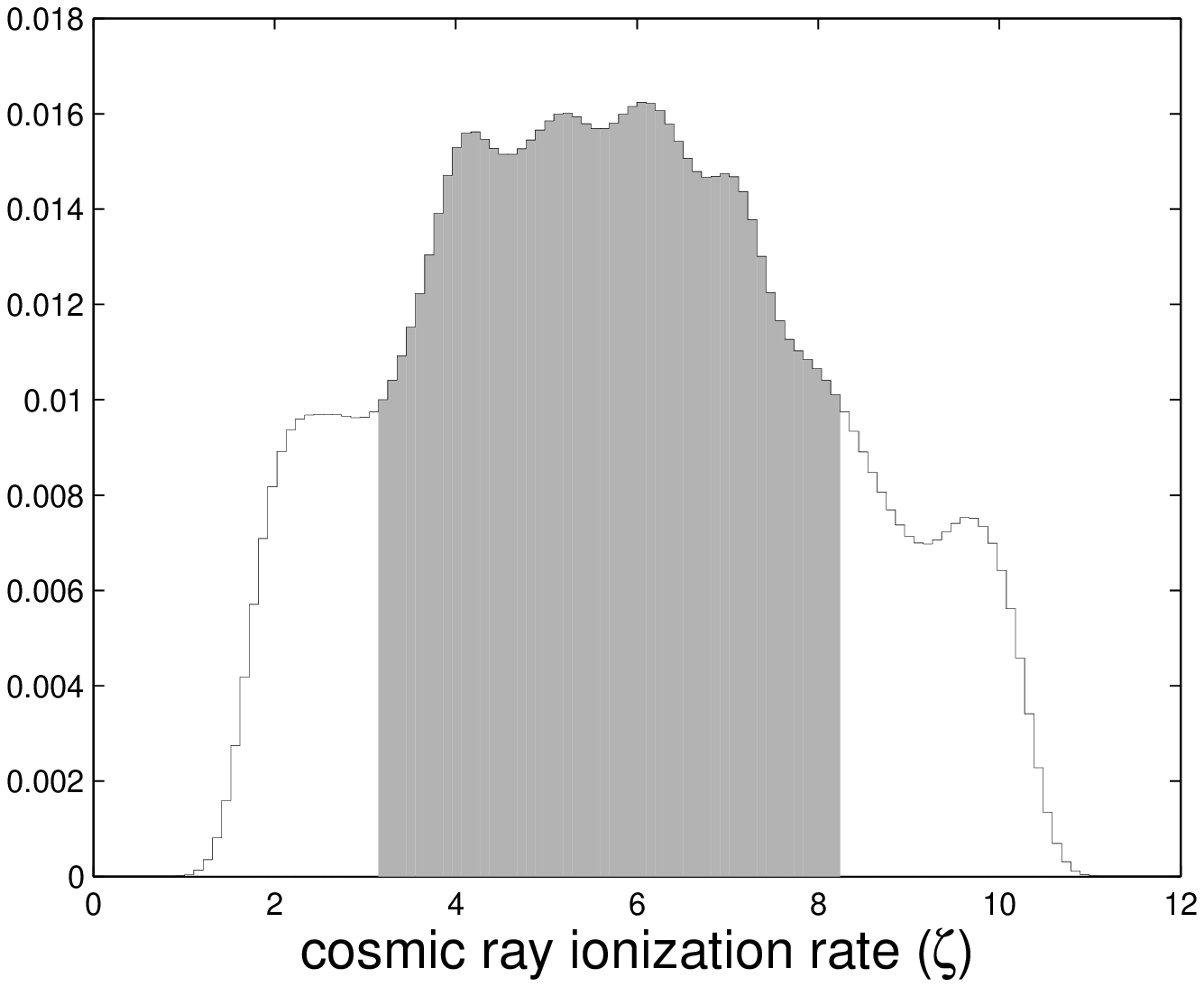}
}
\subfloat[]{
  \includegraphics[width=0.33\textwidth]{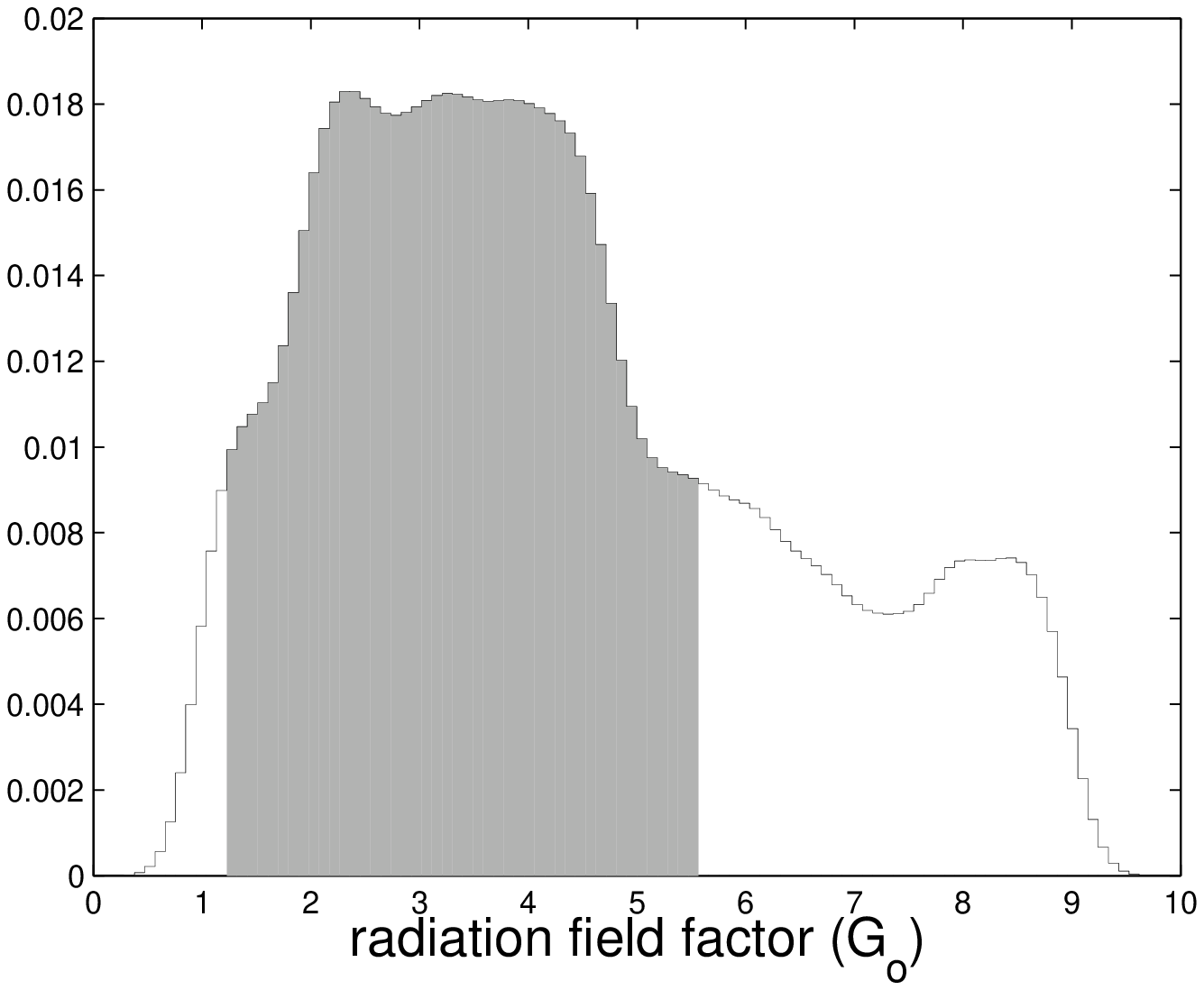}
}
\subfloat[]{
  \includegraphics[width=0.34\textwidth]{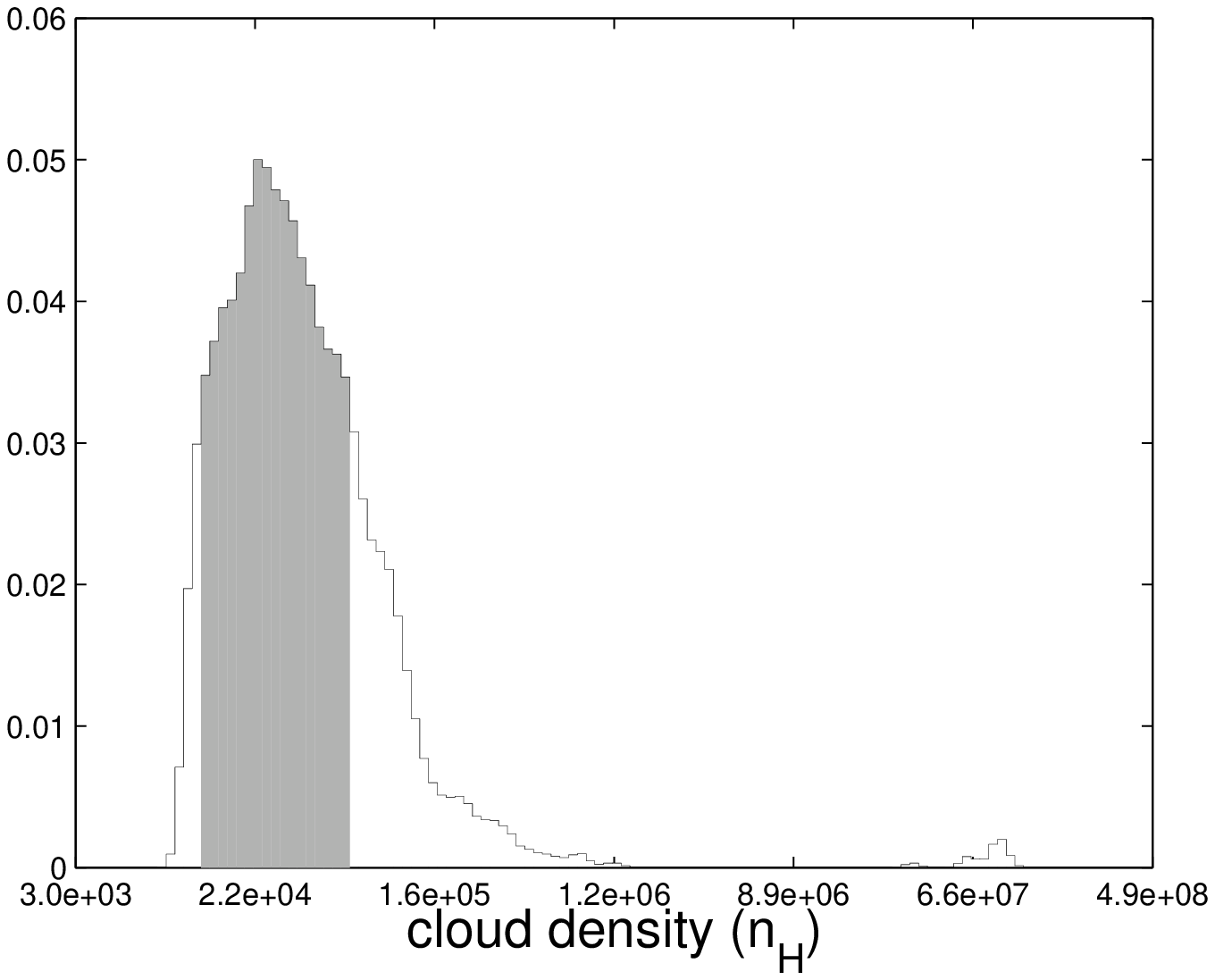}
}
\hspace{0mm}
\subfloat[]{
  \includegraphics[width=0.33\textwidth]{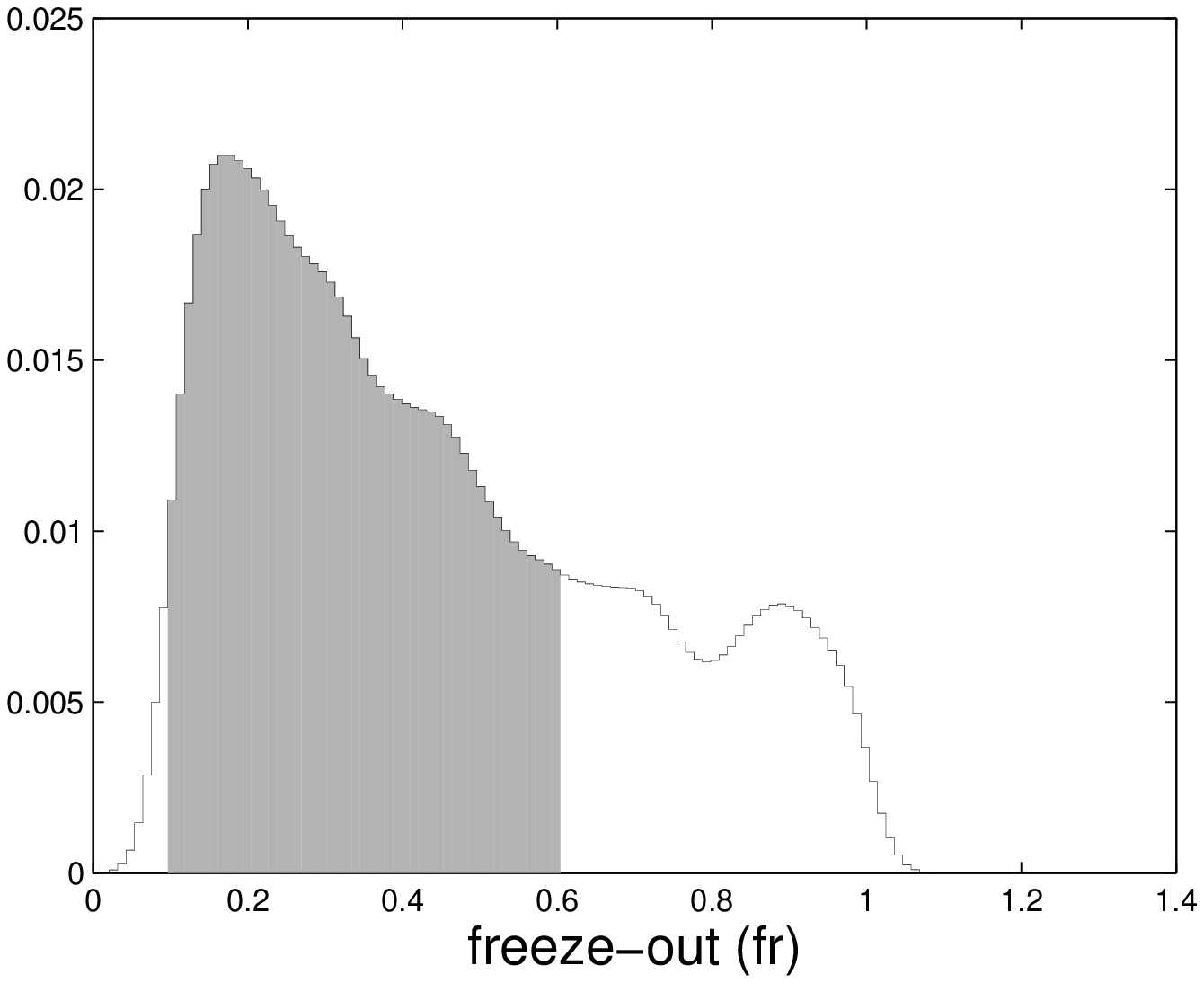}
}
\subfloat[]{
  \includegraphics[width=0.33\textwidth]{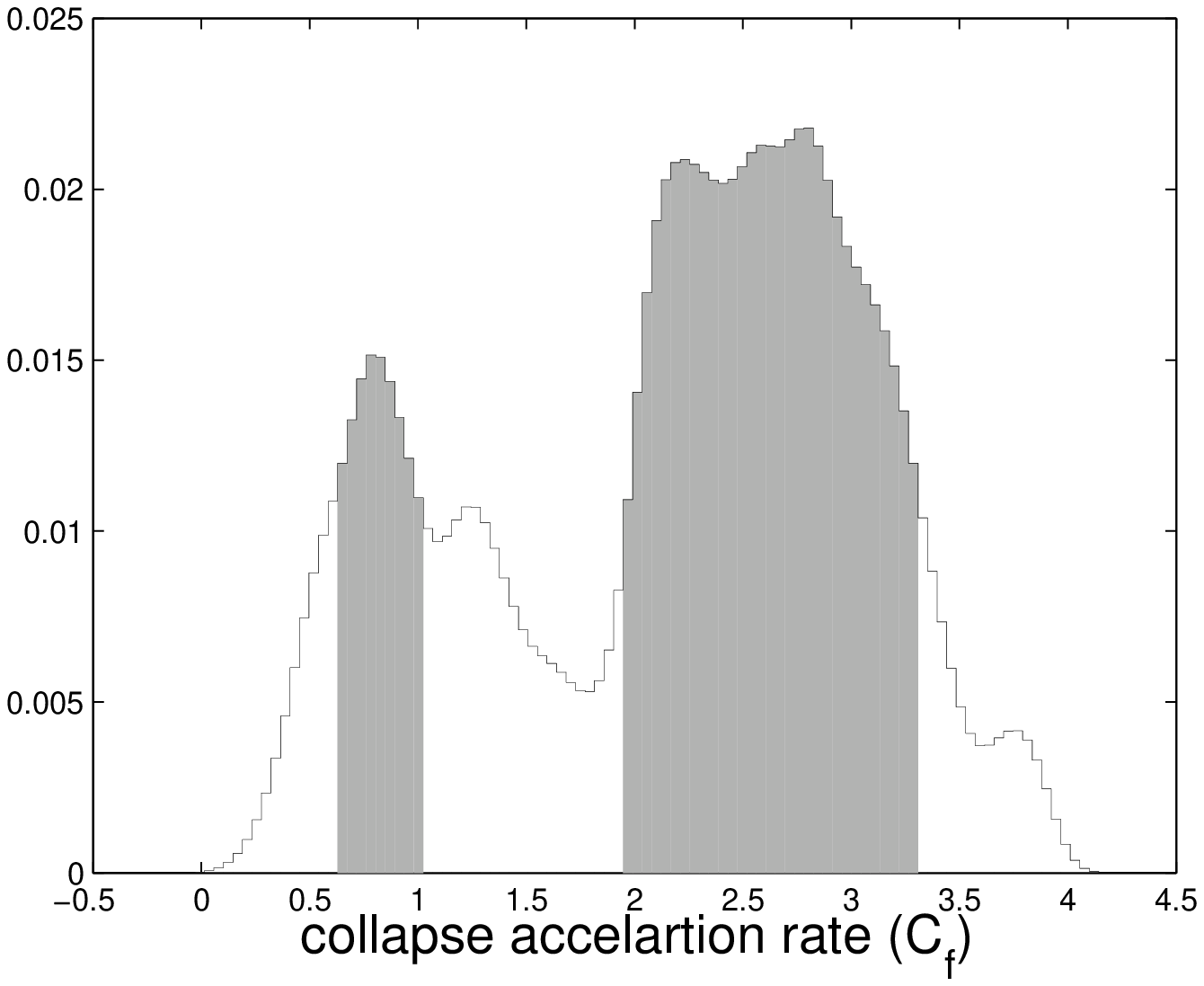}
}
\subfloat[]{
  \includegraphics[width=0.33\textwidth]{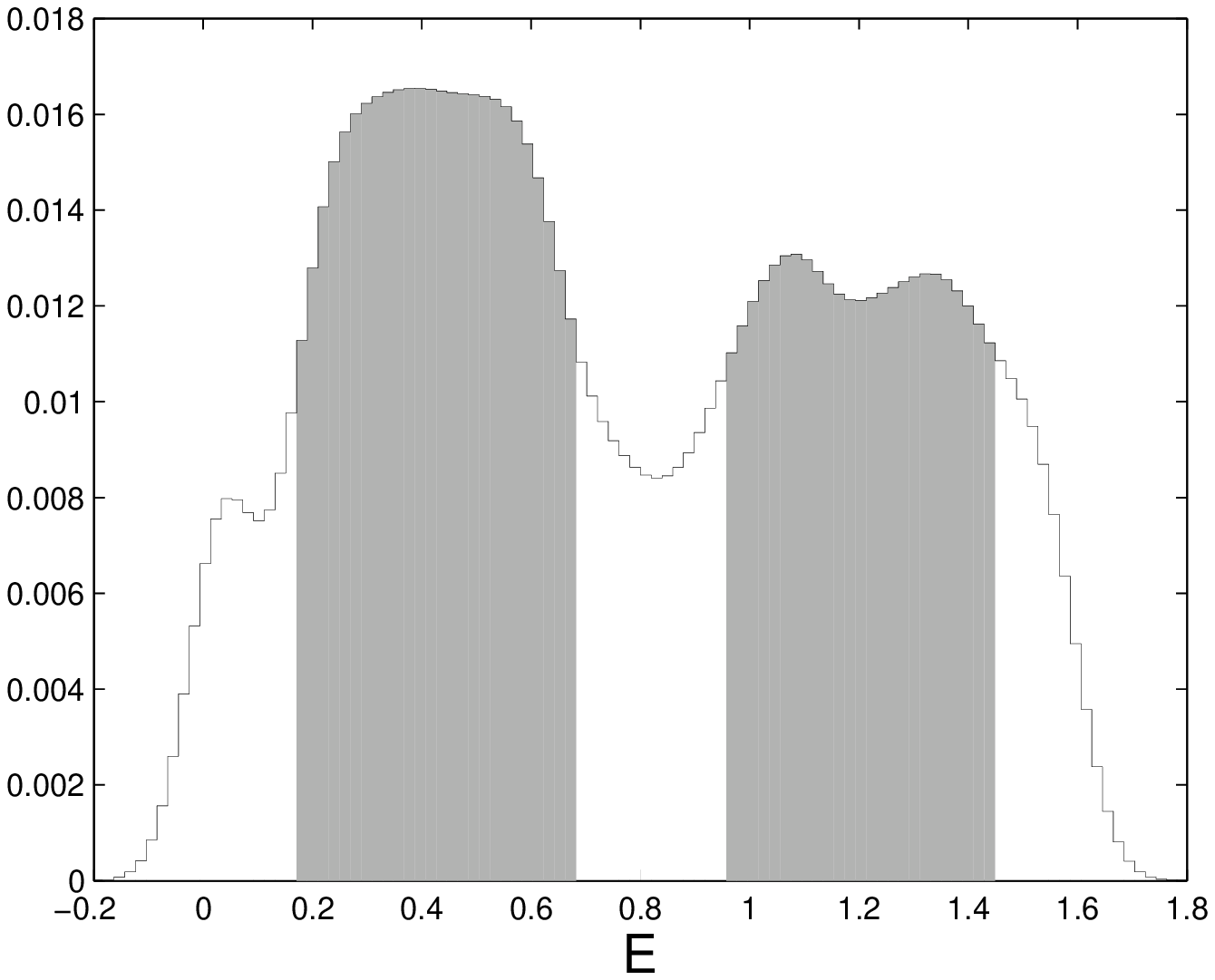}
}
\hspace{0mm}
\subfloat[]{
  \includegraphics[width=0.33\textwidth]{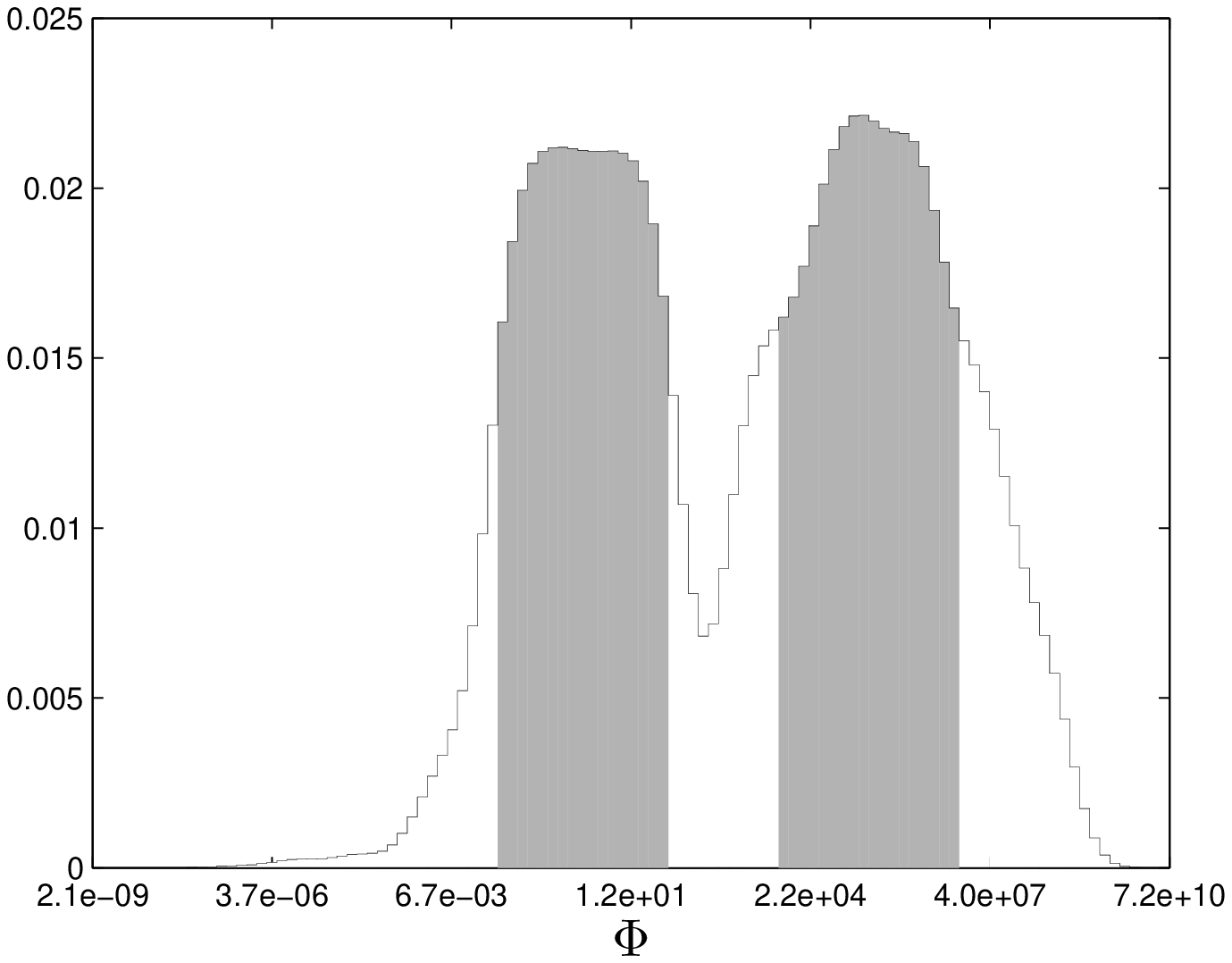}
}
\subfloat[]{
  \includegraphics[width=0.33\textwidth]{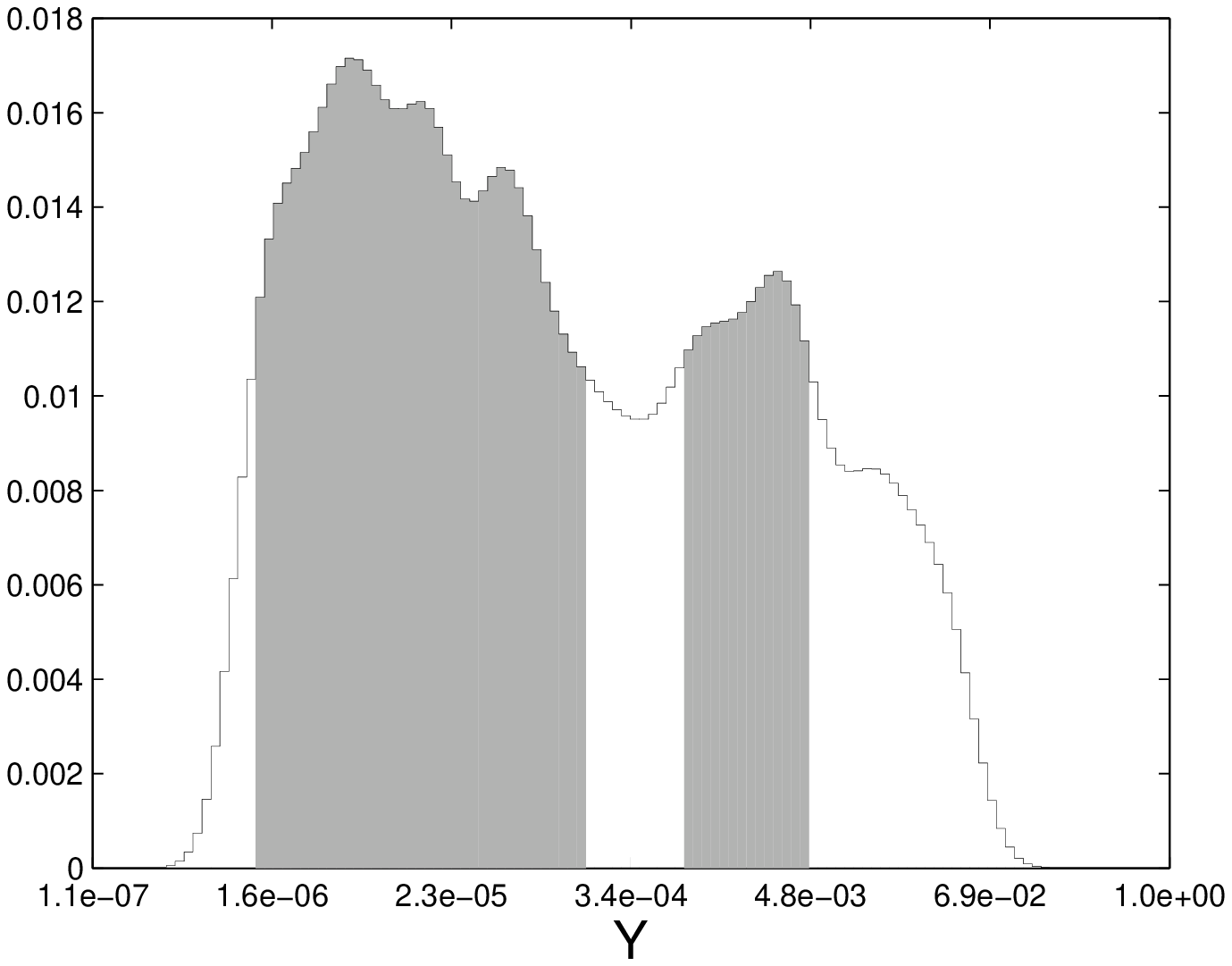}
}
\subfloat[]{
  \includegraphics[width=0.32\textwidth]{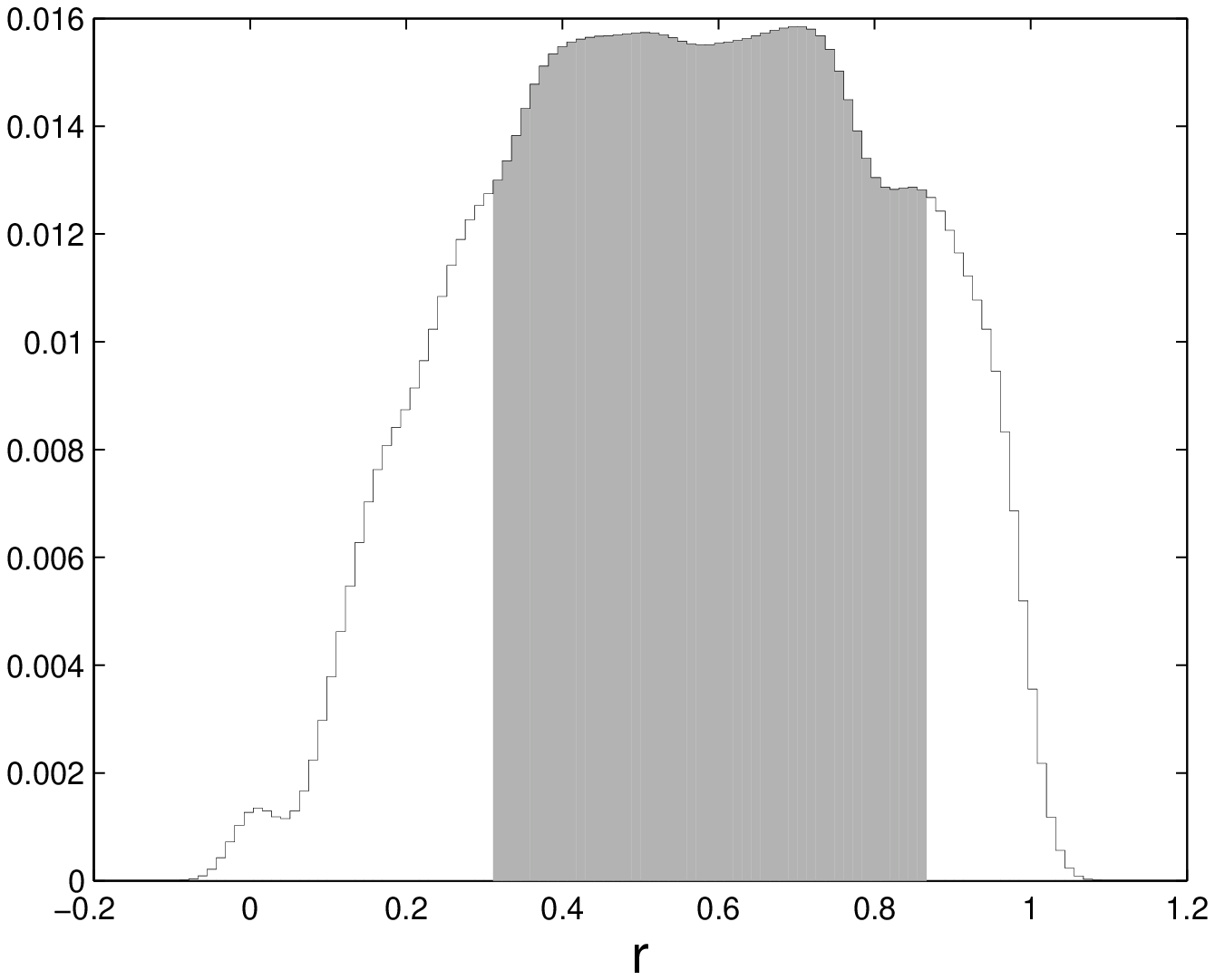}
}
\caption{1D Marginalized PPD for each of the nine parameters using uniform non informative prior. The plots show the Gaussian kernel density estimator of each Probability
Density Function.Darker regions indicate $68\%$ HDR.\label{fig-1}}
\end{figure}

\begin{figure}
\centering
\subfloat[]{
  \includegraphics[width=0.33\textwidth]{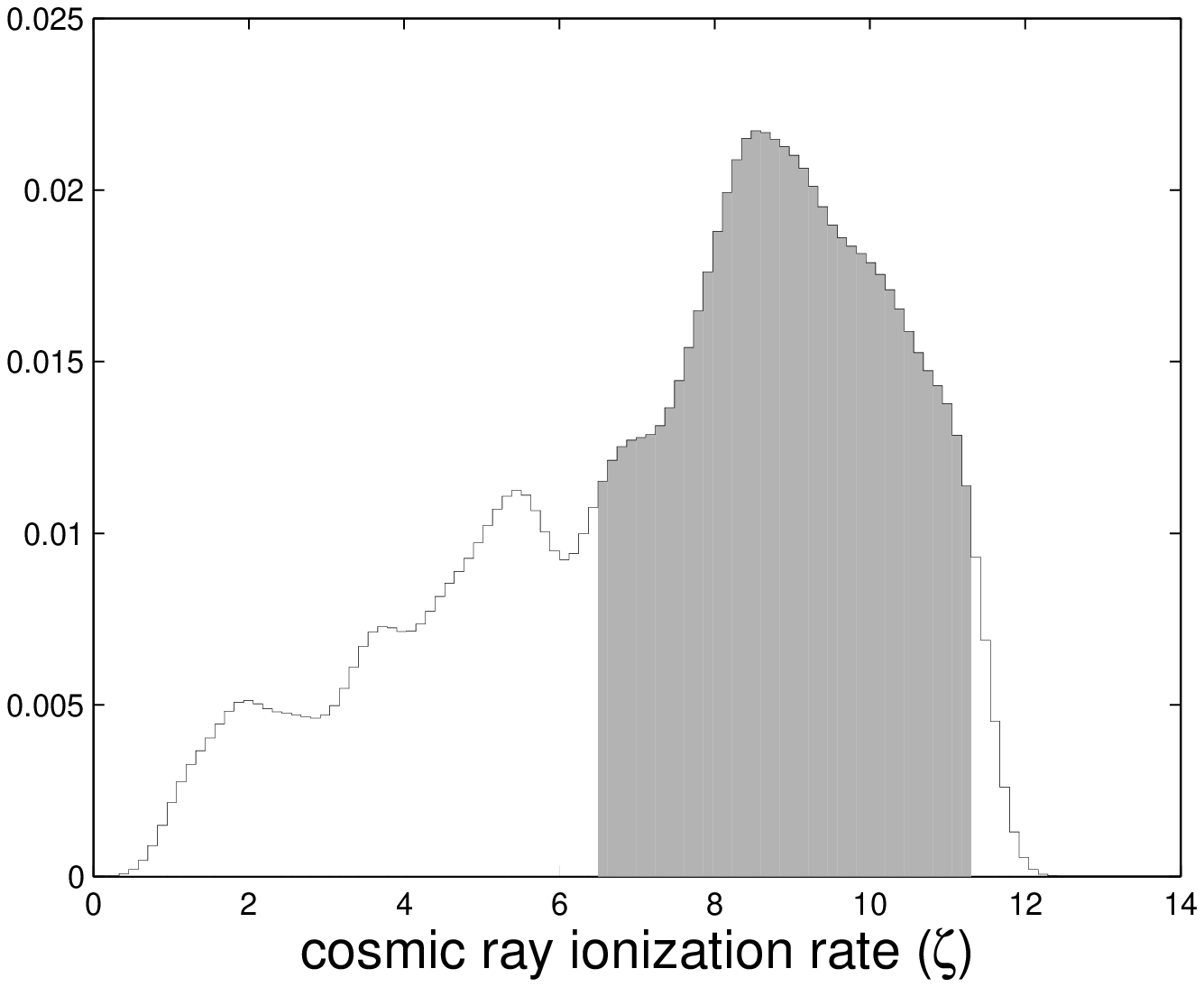}
}
\subfloat[]{
  \includegraphics[width=0.33\textwidth]{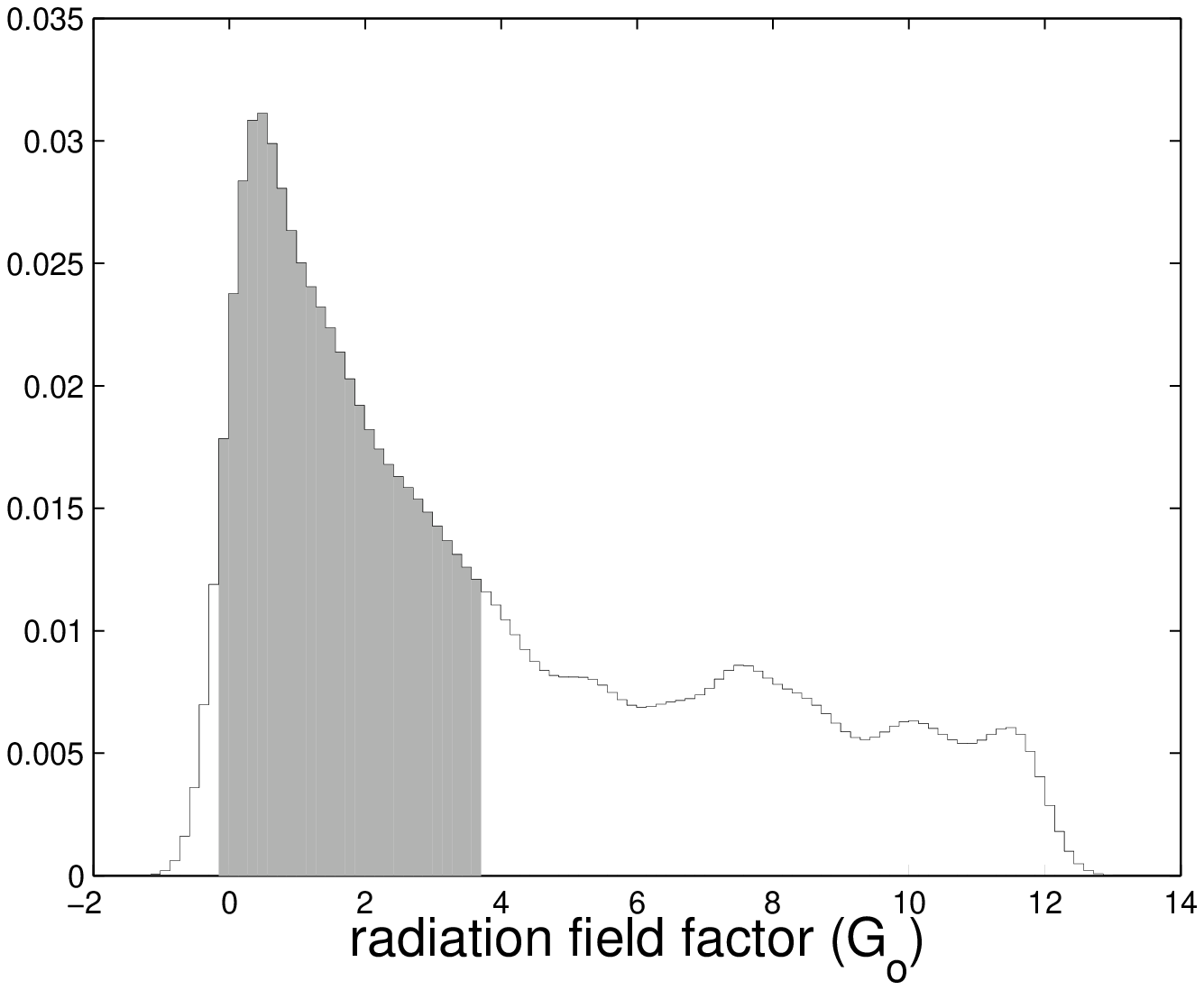}
}
\subfloat[]{
  \includegraphics[width=0.34\textwidth]{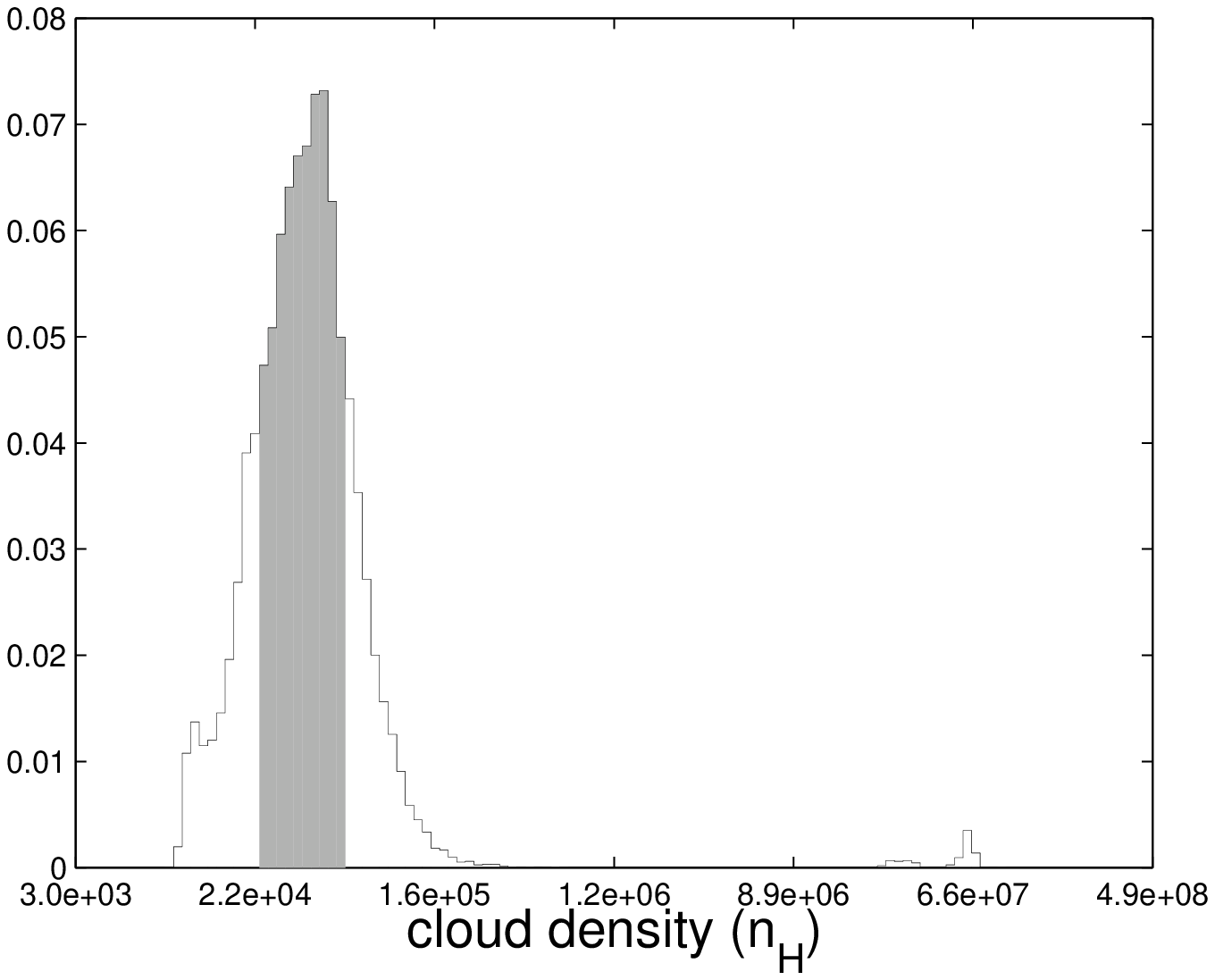}
}
\hspace{0mm}
\subfloat[]{
  \includegraphics[width=0.33\textwidth]{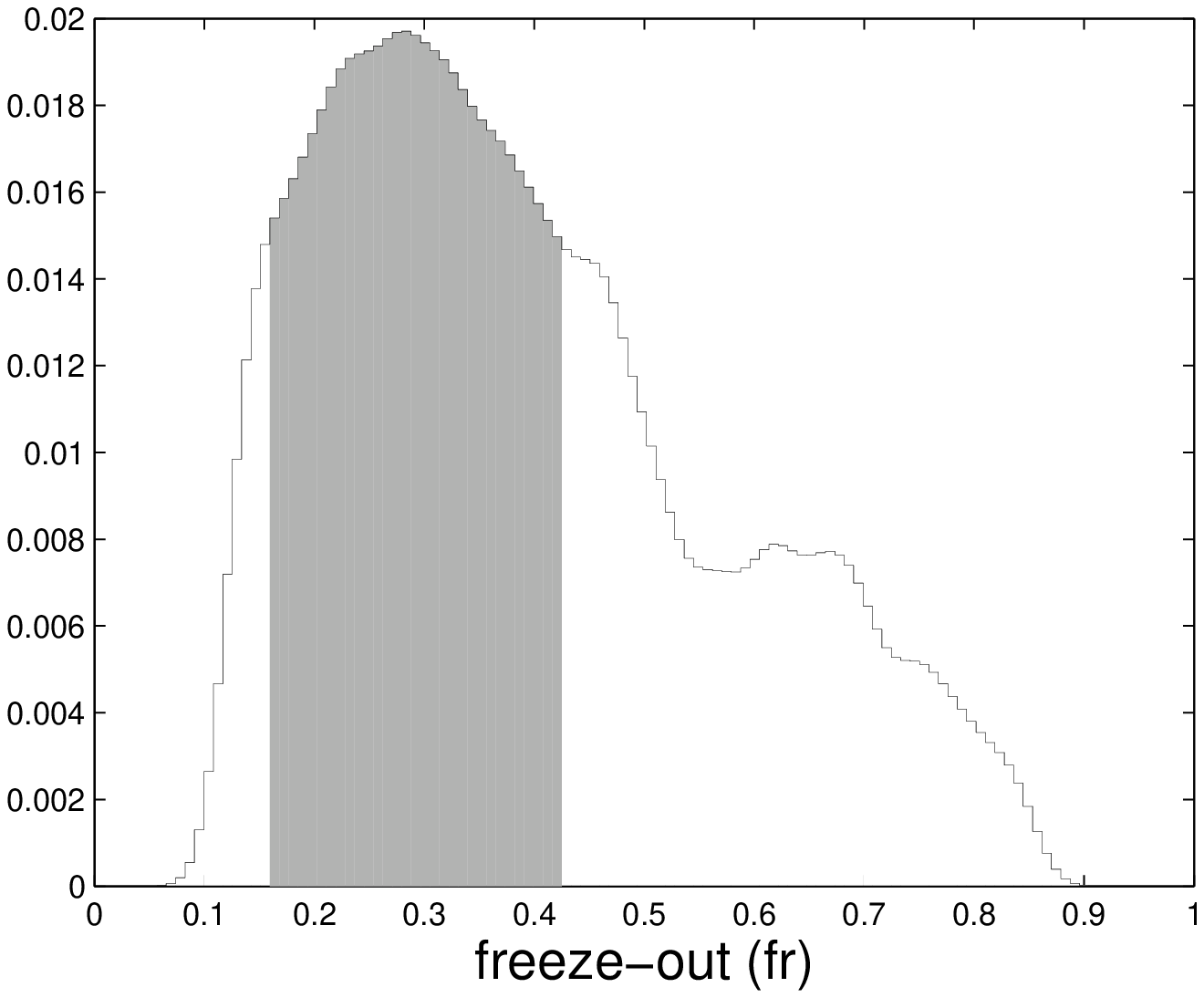}
}
\subfloat[]{
  \includegraphics[width=0.33\textwidth]{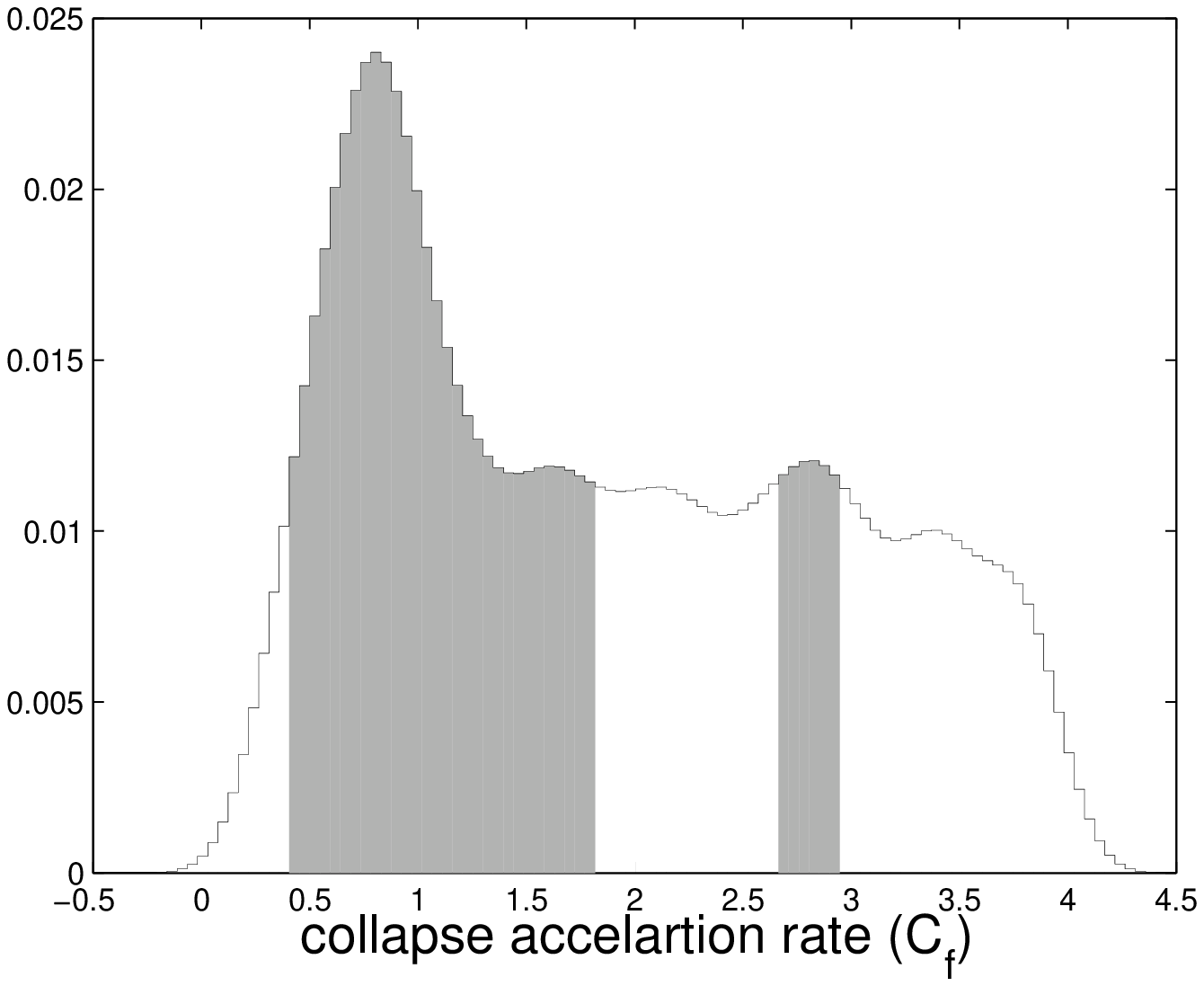}
}
\subfloat[]{
  \includegraphics[width=0.33\textwidth]{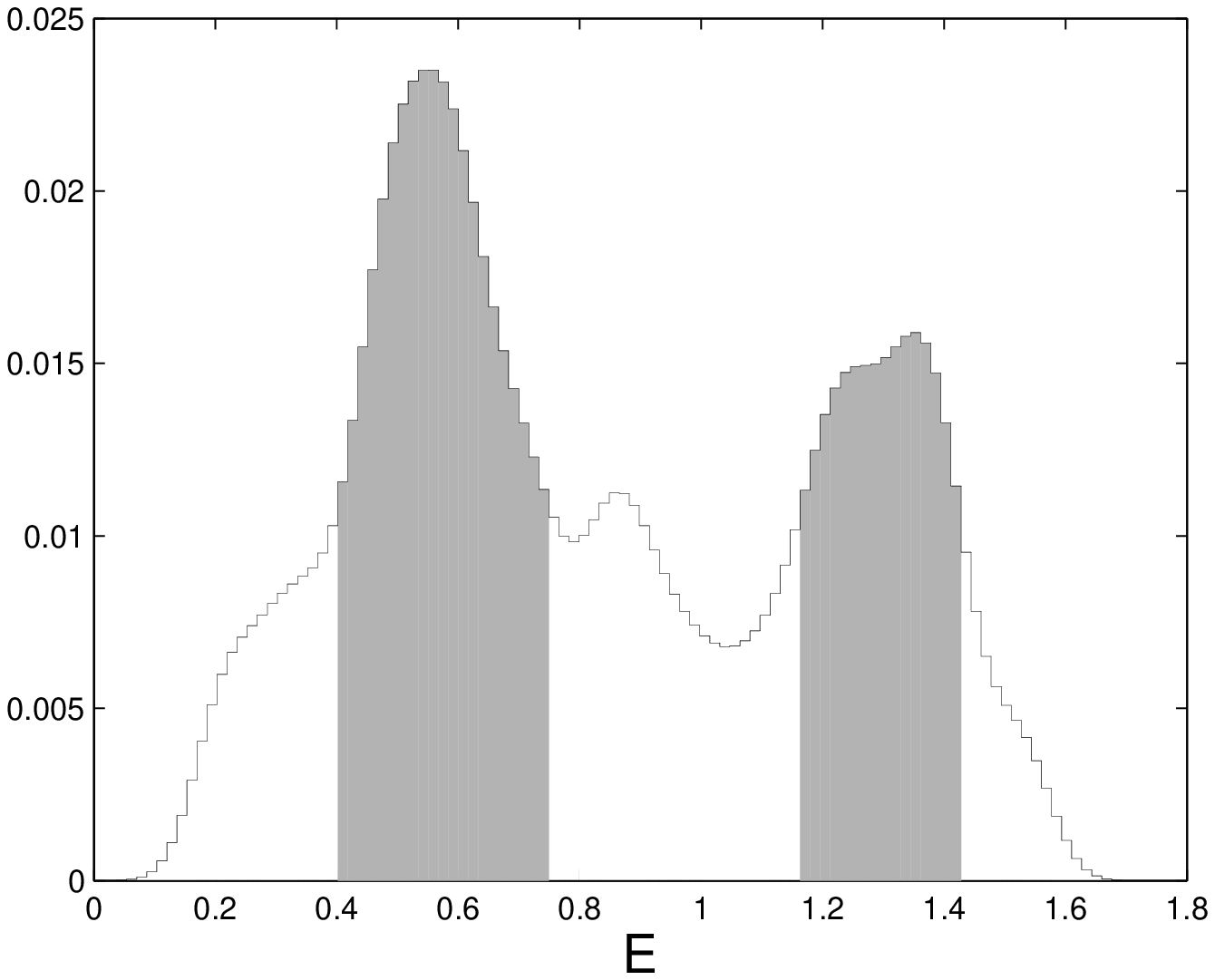}
}
\hspace{0mm}
\subfloat[]{
  \includegraphics[width=0.33\textwidth]{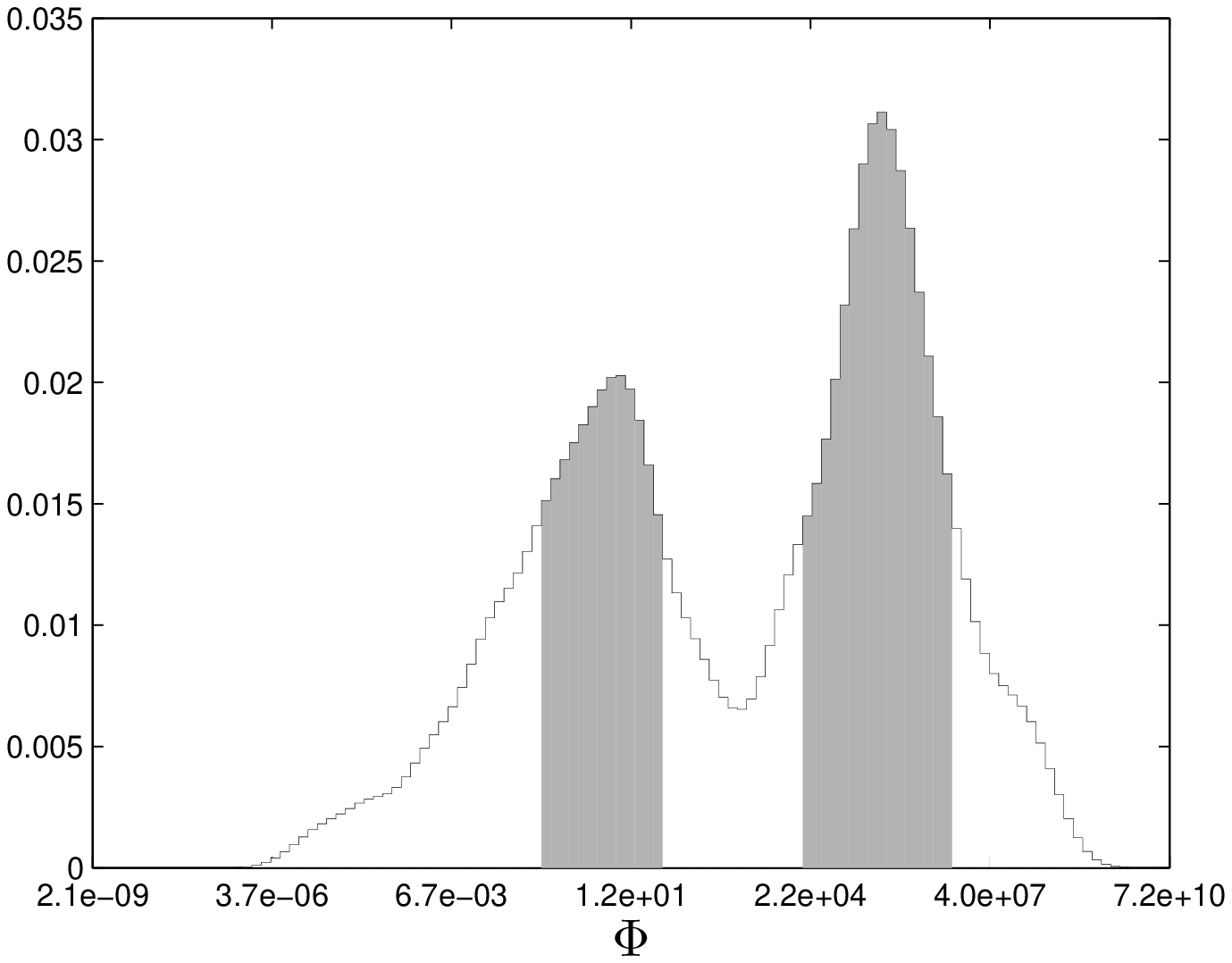}
}
\subfloat[]{
  \includegraphics[width=0.33\textwidth]{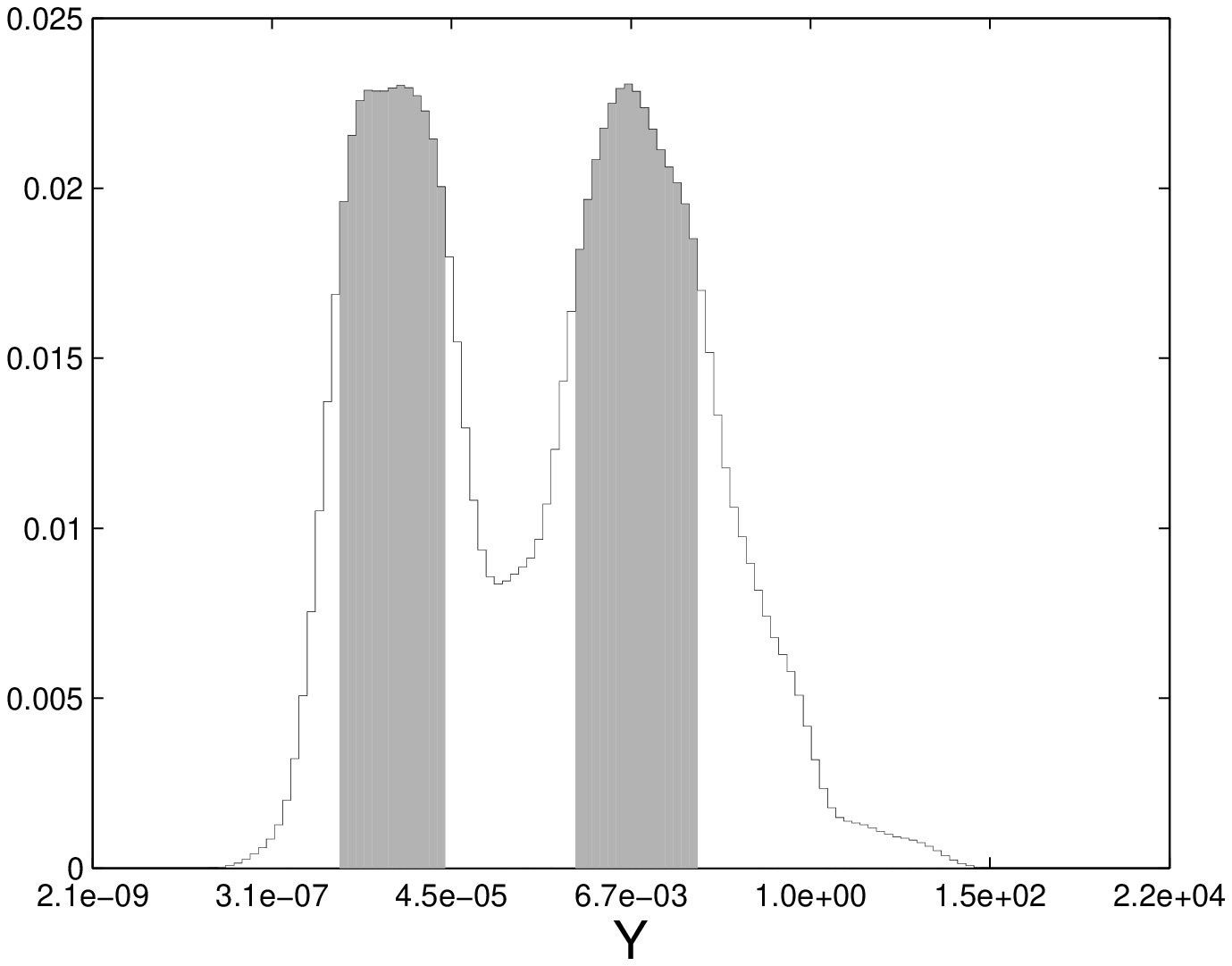}
}
\subfloat[]{
  \includegraphics[width=0.32\textwidth]{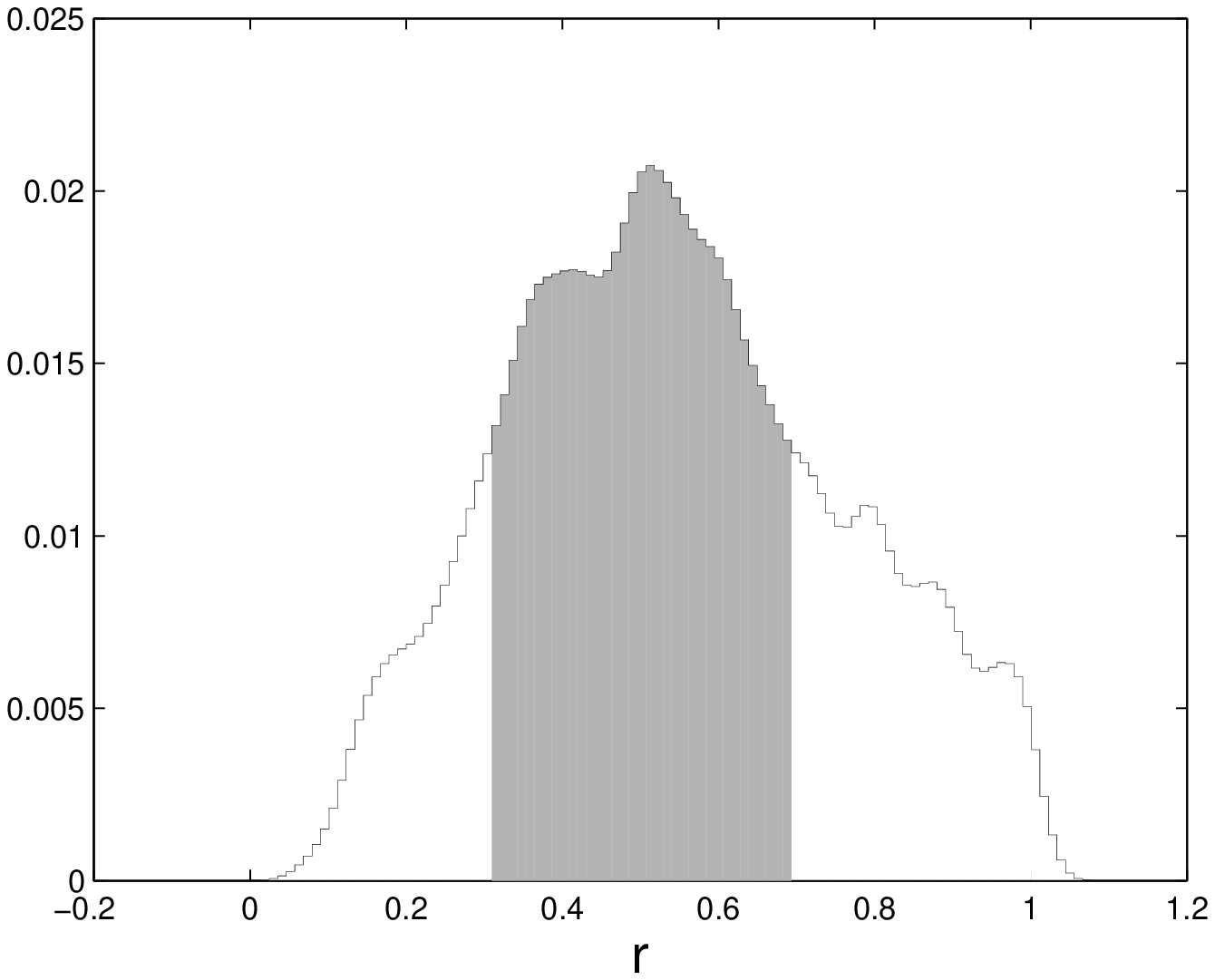}
}
\caption{1D Marginalized PPD for each of the nine parameters using informative prior from gas phase chemistry. The plots show the Gaussian kernel density estimator of each Probability
Density Function. Darker regions indicate $68\%$ HDR. \label{fig-2}}
\end{figure}

When quoting parameter estimation results and especially multivariate results, it is convenient to decrease the parameter space to posterior intervals 
about single marginalized parameters.
Figure ~\ref{fig-x1} shows the nine 1D marginalized posterior probability distributions 
of the parameters for the benchmark test using a uniform prior. 
In Figure~\ref{fig-1}, we present the nine 1D marginalized posterior probability distributions 
of the parameters and their 68\% High Density Regions (HDR), recovered from the uniform prior case. Figure~\ref{fig-2}, presents the same results 
for the informative prior case. HDR indicate the parameter space where the probability density is higher. We refer readers seeking more details 
about marginal posterior probability function and High Density Regions to Appendix C and references therein. In order to compare the 2 prior 
cases and quantify the level of constraint for each parameter we introduce a 
measure of parameter constrain, the High Density Spread (HDS), which is defined as follows:

Let $|HDR|$ be the width of a High Density Region of a parameter's k  density function with definition domain $\mathbb{D}_{{\theta_k}}$ and $|\mathbb{D}_{{\theta_k}}|$ the width of 
the domain. Width is defined with respect to some simple measure such as the Lebesque measure \citep{Leb1902}. Then the High Density Spread is defined as :
\begin{displaymath}
 HDS = \frac{|HDR|}{|\mathbb{D}_{{\theta_k}}|}
\end{displaymath}
The HDS ratio can be perceived as an index of the level of uncertainty on a predefined definition domain and the higher it is the less constrained is a 
parameter. Table~\ref{tbl-3} presents
HDS for each parameter for both priors used. Figure~\ref{fig-3} shows the 2 dimensional marginal PPD for parameters that present statistical interest. Finally, 
Table~\ref{tbl-4} lists the statistical mean and standard deviation for the $\sim 35\%$ HDR of the joint distribution for all the 9 parameter. The general statistical 
picture we get from Figures~\ref{fig-1} and~\ref{fig-2}
shows that the distributions of all the parameters are far from Gaussian and most of them have more than one modes. Looking at the models with physical units, 
we can also notice that most of the density lies away from the limits of our definition domain for both cases, which validates our choice for $\mathbb{D}_{{\theta}}$.

\subsection{Blind Benchmark Test Results}
\label{sec-3.x1}
\begin{figure}
\centering
\subfloat[]{
  \includegraphics[width=0.33\textwidth]{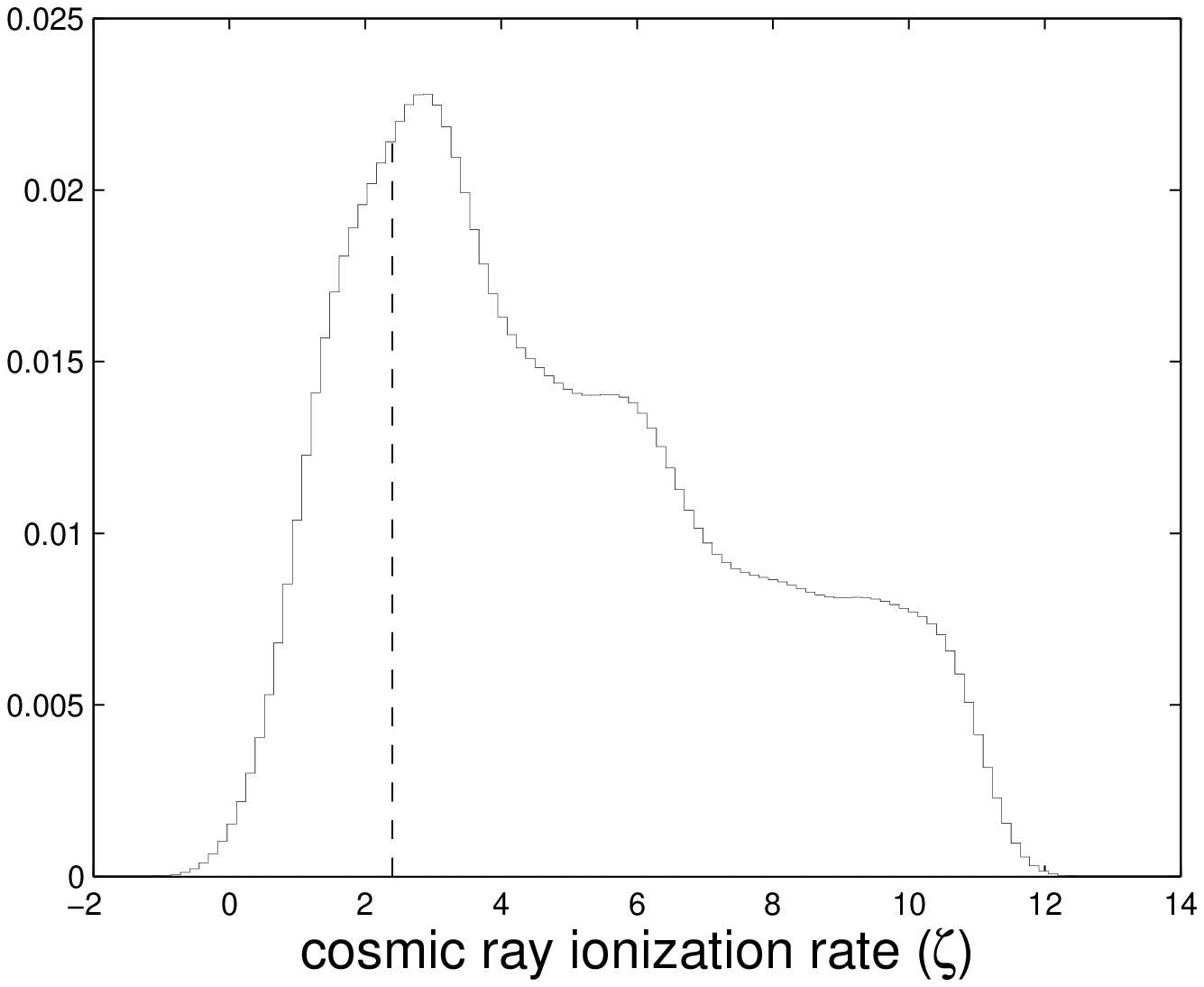}
}
\subfloat[]{
  \includegraphics[width=0.33\textwidth]{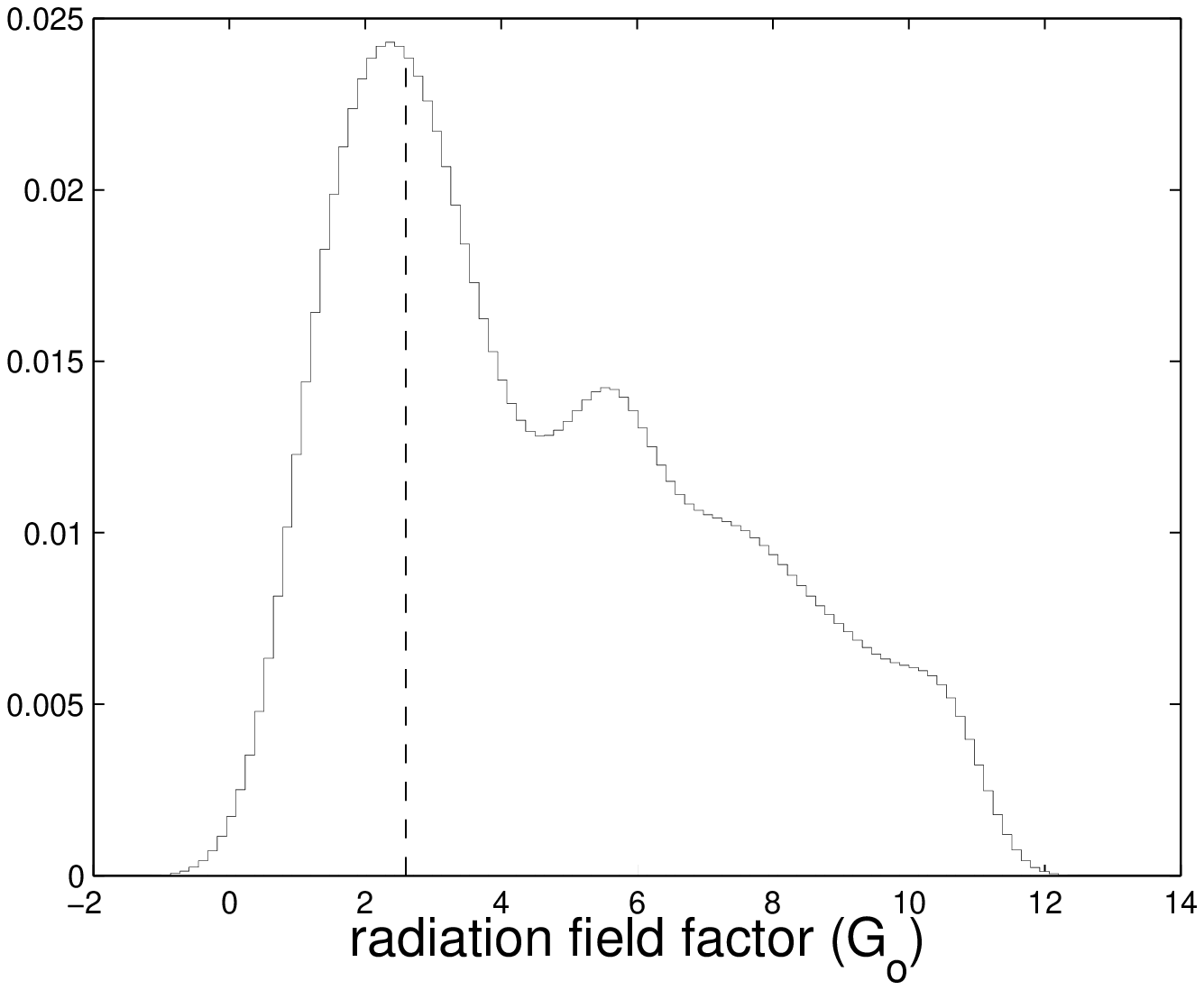}
}
\subfloat[]{
  \includegraphics[width=0.34\textwidth]{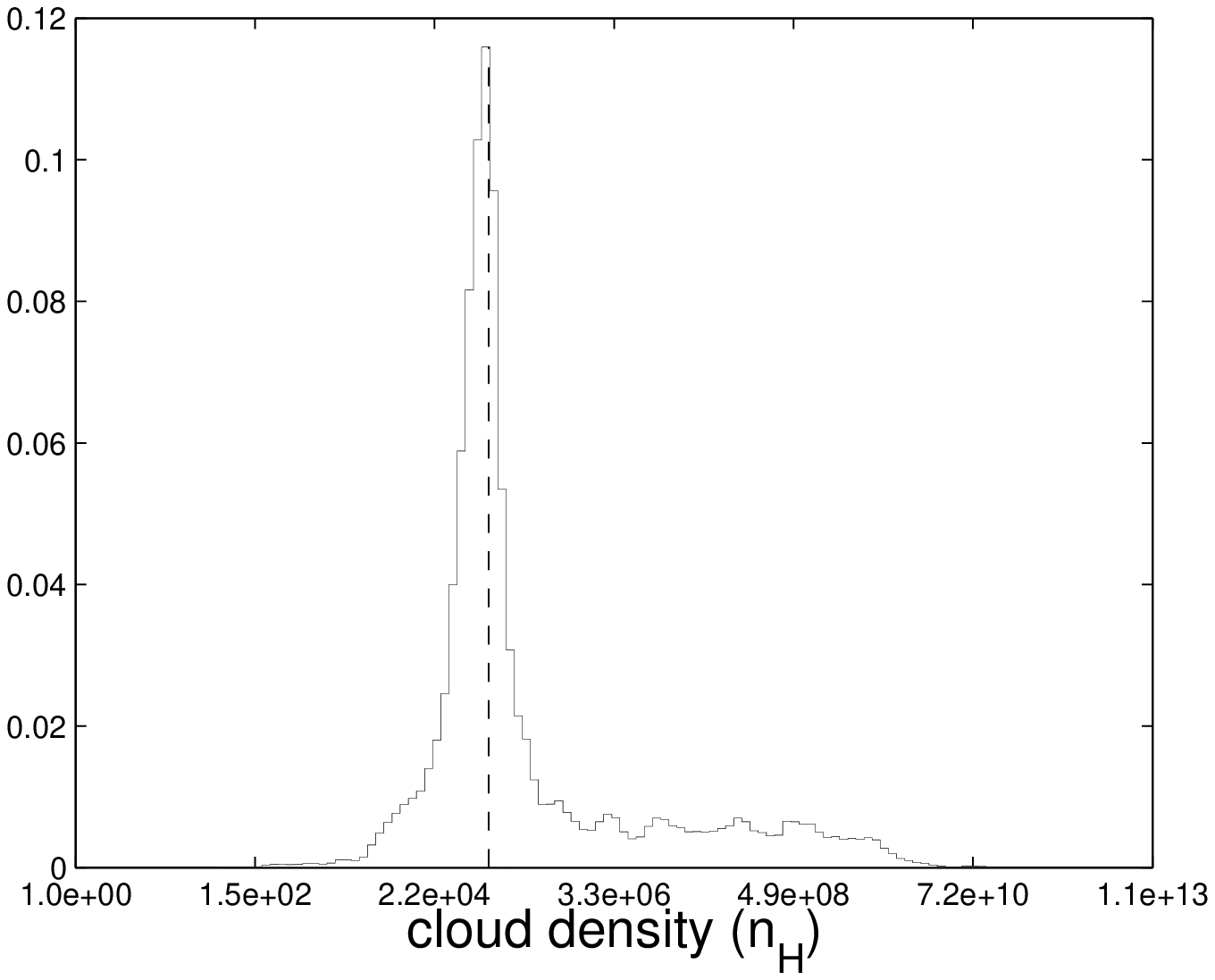}
}
\hspace{0mm}
\subfloat[]{
  \includegraphics[width=0.33\textwidth]{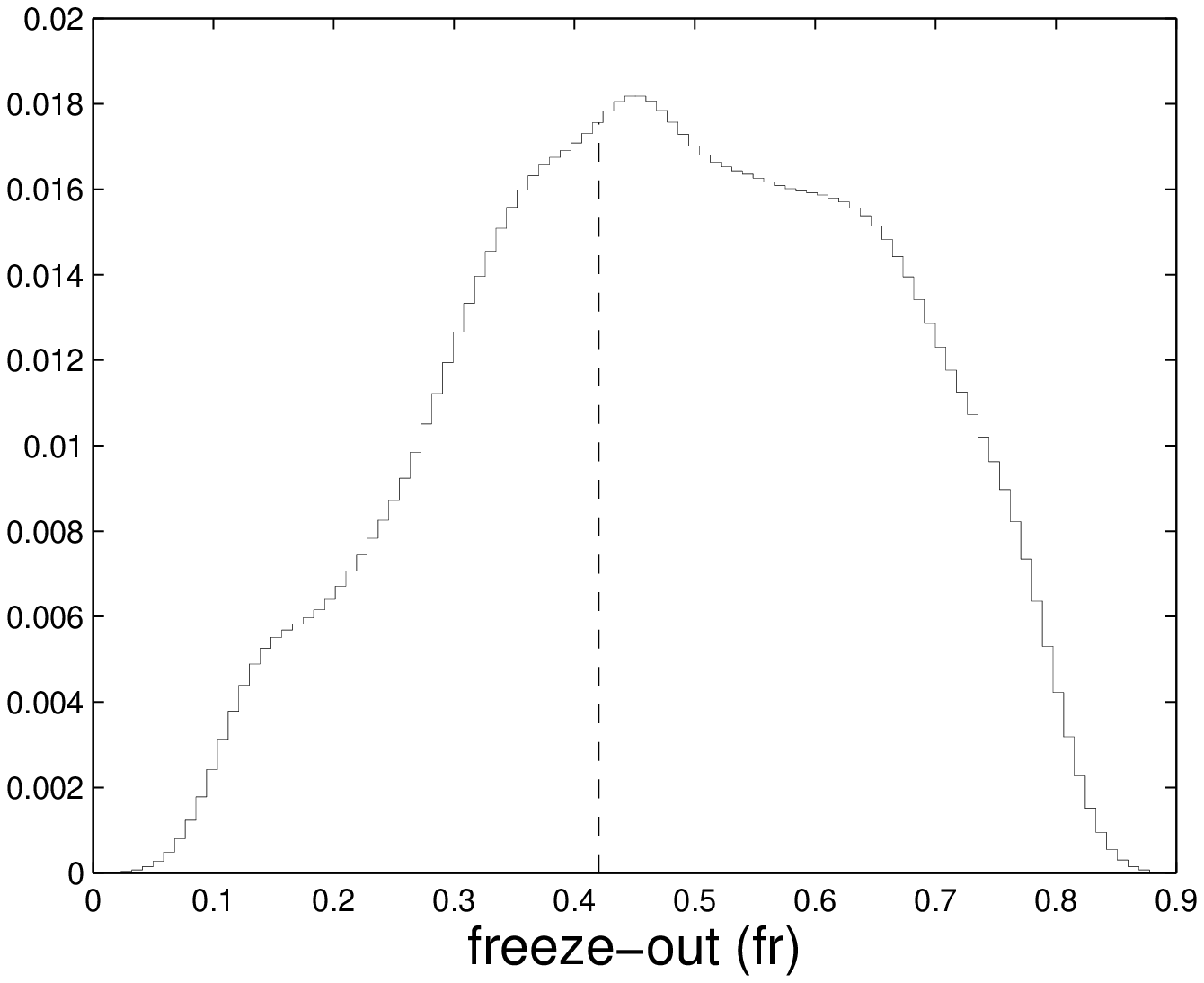}
}
\subfloat[]{
  \includegraphics[width=0.33\textwidth]{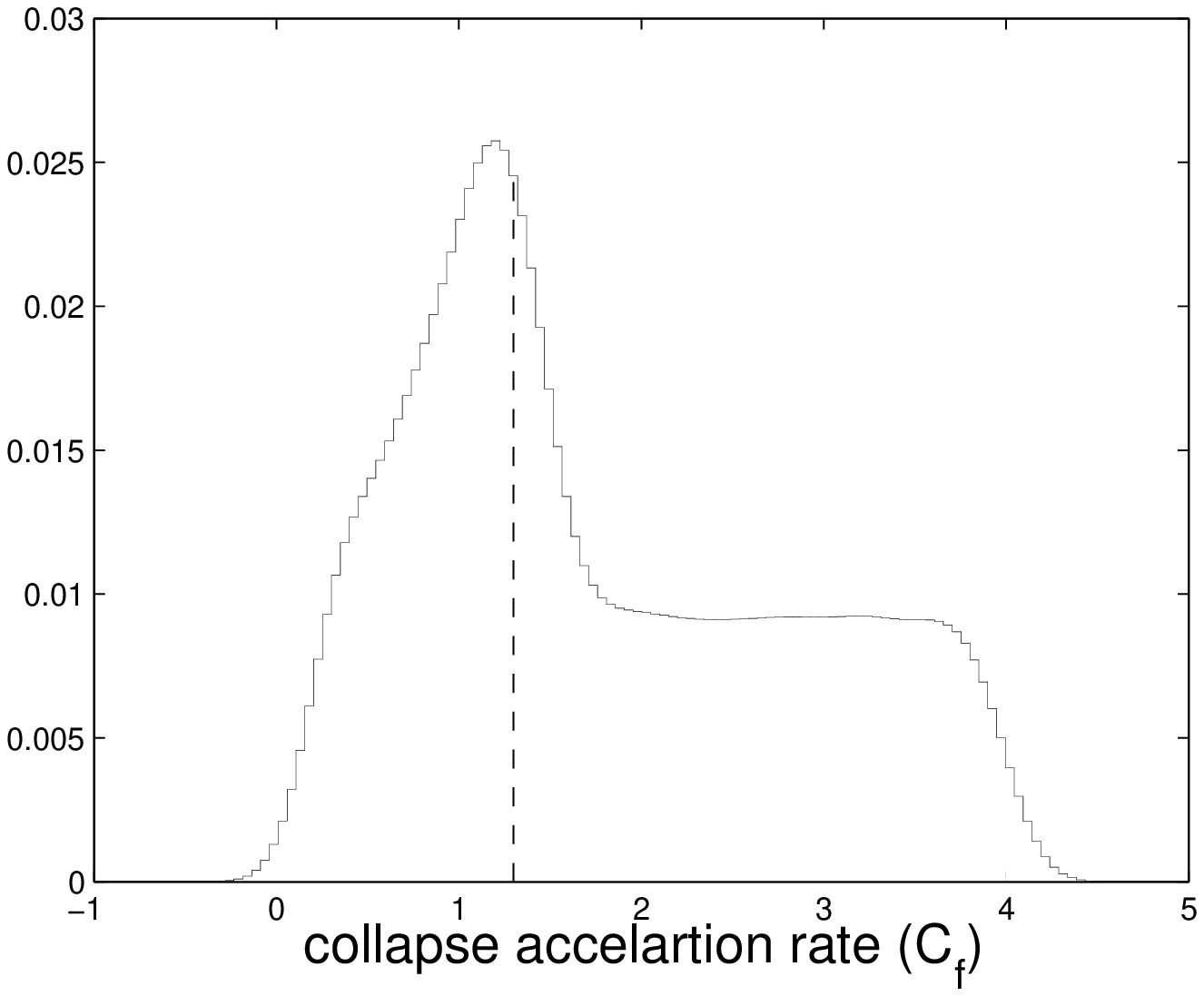}
}
\subfloat[]{
  \includegraphics[width=0.33\textwidth]{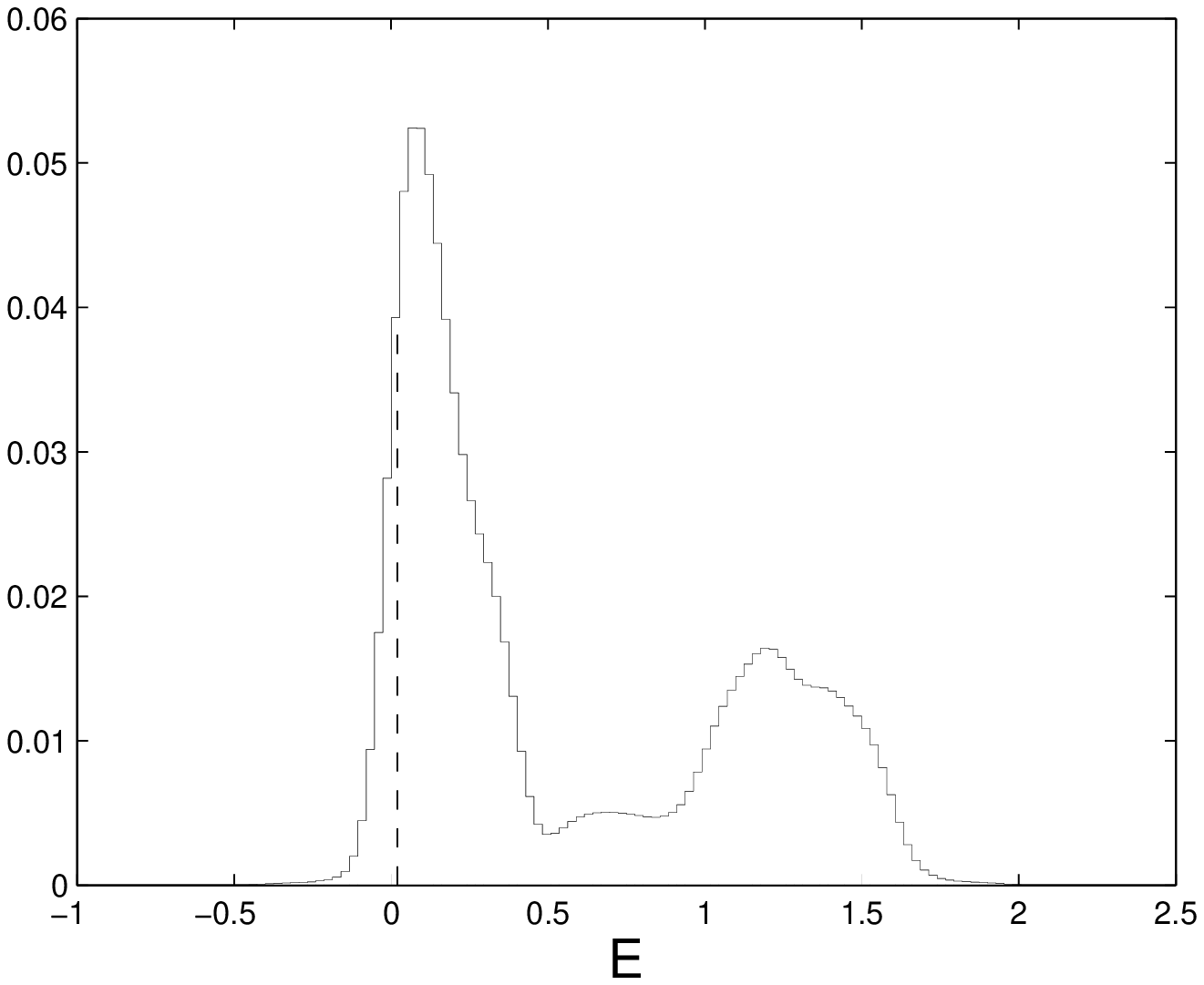}
}
\hspace{0mm}
\subfloat[]{
  \includegraphics[width=0.33\textwidth]{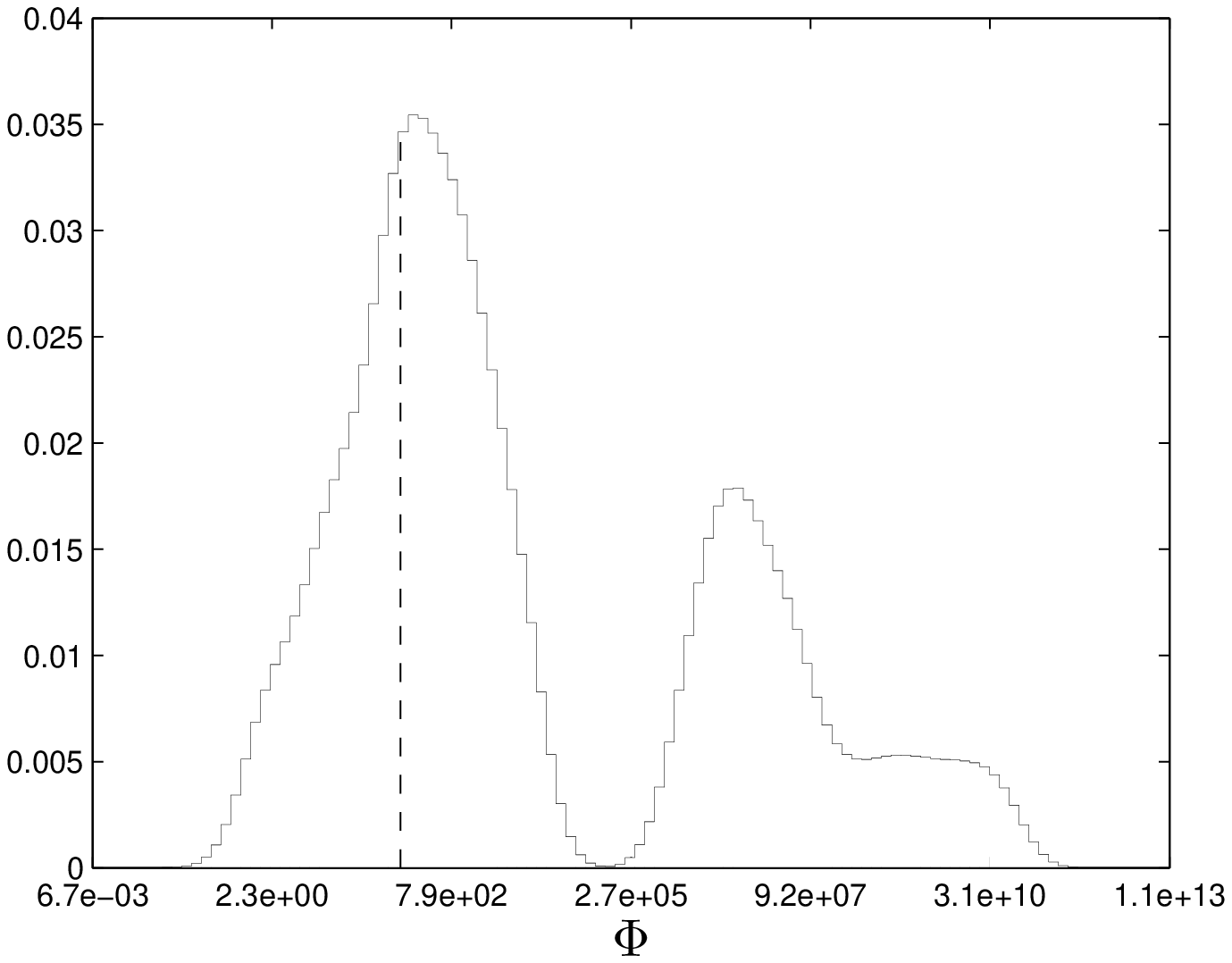}
}
\subfloat[]{
  \includegraphics[width=0.33\textwidth]{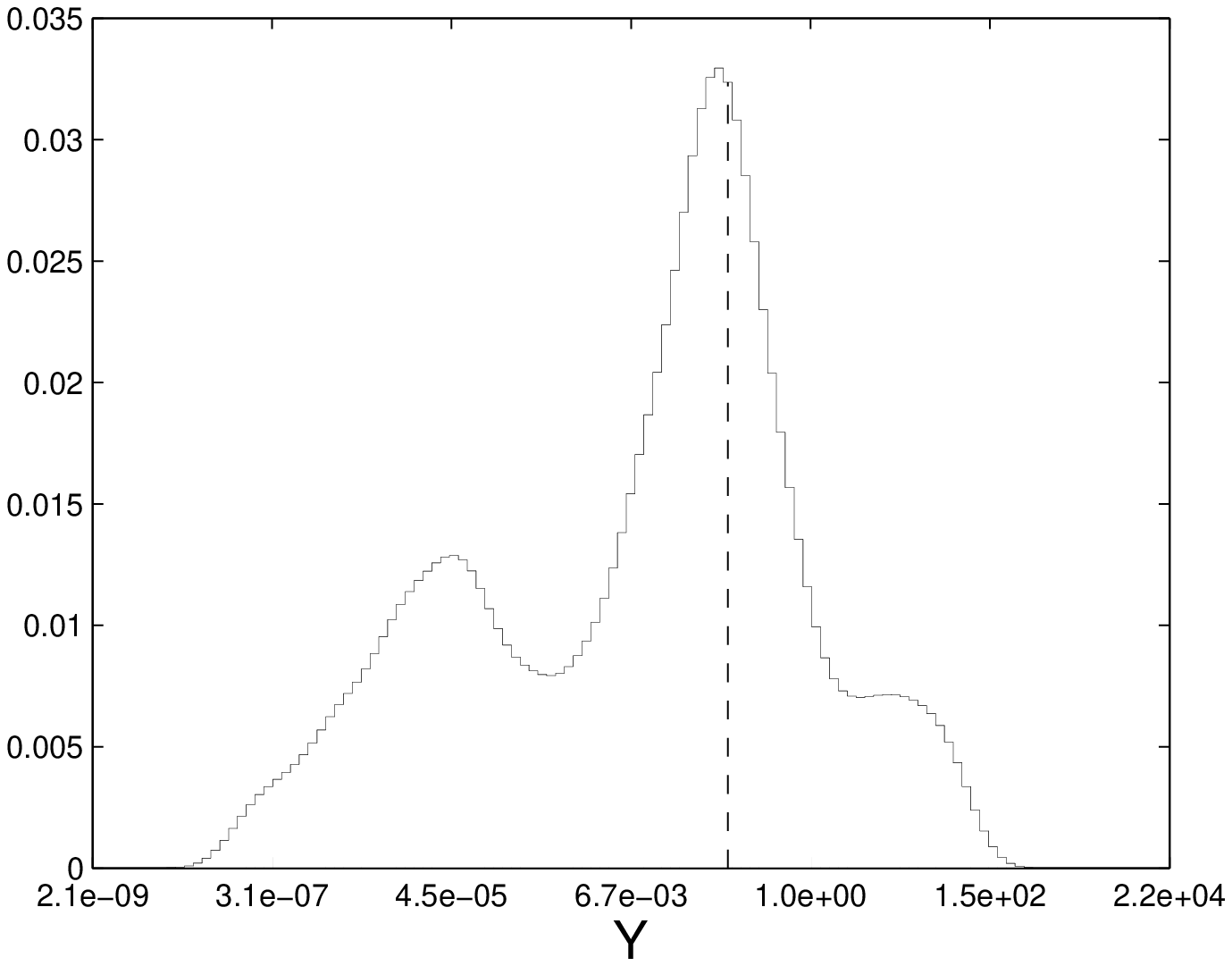}
}
\subfloat[]{
  \includegraphics[width=0.32\textwidth]{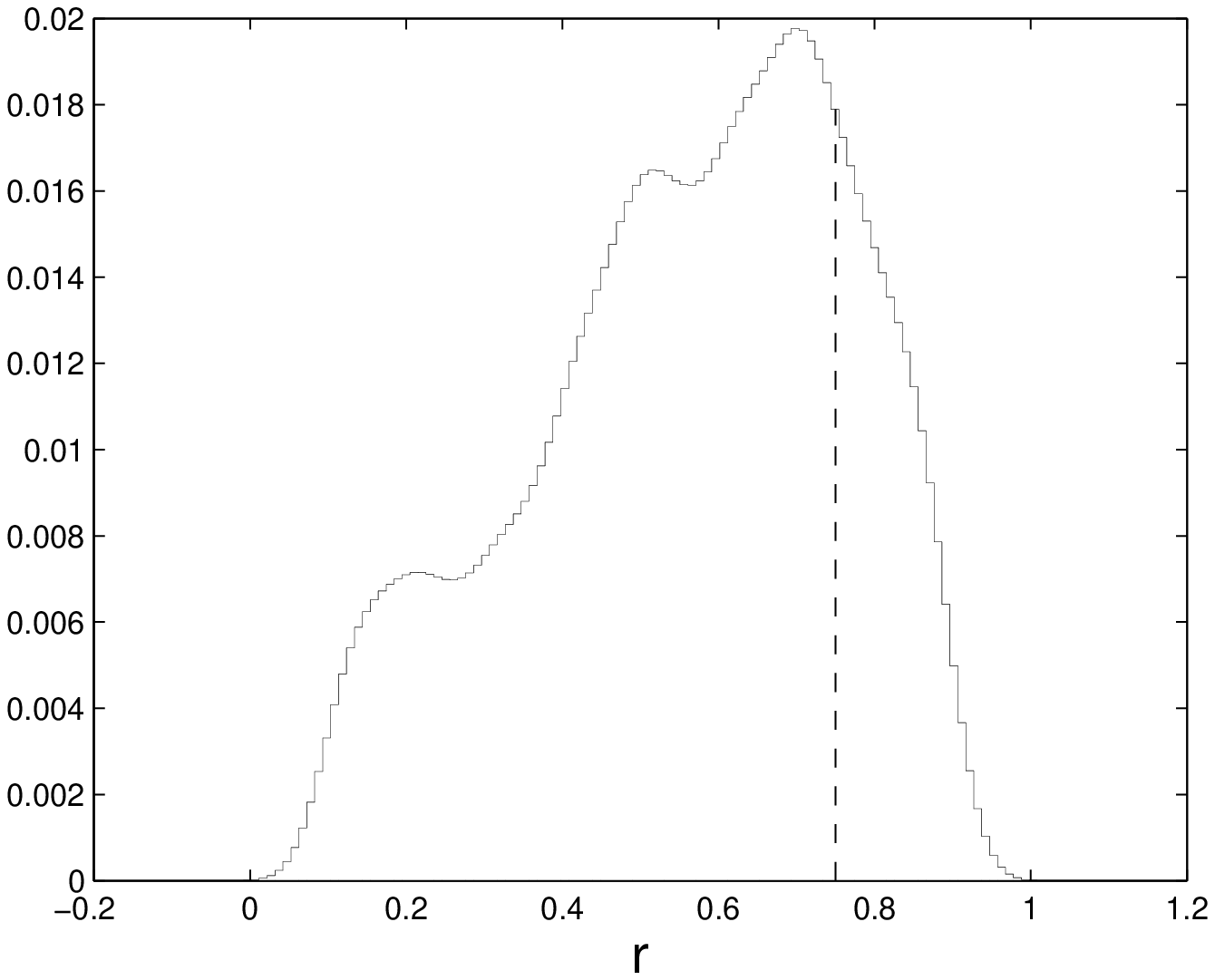}
}
\caption{1D Marginalized PPD for each of the nine parameters for the Blind Benchmark Test. The plots show the Gaussian kernel density estimator of each Probability
Density Function. Dashed lines indicate the pre-defined parameter values $\boldsymbol{\theta_T}$ we wish to recover.\label{fig-x1}}
\end{figure}

The results of the performed test as shown in Figure~\ref{fig-x1} reveal two important insights. First of all, high probability density regions for all the 
parameters include and hence recover the true parameters. As we can see in Figure~\ref{fig-x1}, all the pre-defined parameter values lie under or very close to 
the highest density point of the marginal PPD.
This result simply validates that both the Bayesian approach makes accurate inference based on the
given observations and the MH algorithm samples efficiently the solution space. Secondly, we can observe that in many cases there are additional high probability
density regions. These regions prove and highlight the ill-posed nature of our problem by indicating that different parameter 
sets can produce similar observations. Combining the two insights, we can conclude that 
the Bayesian method with MCMC sampling is exploring efficiently the parameter space, revealing the solution regions that answer our ill-posed inverse problem.
In addition, we can conclude that in order to constrain our solution space we should either introduce numerical regularization factors 
(e.g. gas phase species) or scientific prior knowledge.

\subsection{Influence of priors}

A visual comparison of Figures~\ref{fig-1} and~\ref{fig-2} reveals what we can quantitatively observe in Table~\ref{tbl-3}. With non-informative uniform prior 
the high density regions seem to cover large sections of the distribution, which in some cases reach $50\%$ of the definition domain. This means 
that most of the parameters are not constrained enough. The most statistically straightforward parameters seem to be clearly the $n_H$ and then the
$\phi$ and $fr$ parameters, presenting distinct modes and relatively low HDS. $G_{\circ}$ and $\zeta$ seem neither 
constrained nor relevant enough, while $fr$ seems to have a clear mode, followed by a very heavy tail. The rest of the parameters present high HDS,
above $40\%$ with several disjoint high density regions and do not allow us to reach credible conclusions about the parameters.
Including the prior information from the gas phase species changes the picture significantly as can be seen in both Figure~\ref{fig-2} and Table~\ref{tbl-3}.
We can observe that the HDR get smaller and the parameters seem more constrained. The distribution of $\zeta$ is now denser around high 
values ($>6$), while $G_{\circ}$ has to be low ($<4$). The $n_H$ remains well constrained with even 
lower HDS, while the distribution of $fr$ now clearly constraints the parameter to low domain values. The distribution of $C_f$ is also altered 
significantly: not only the HDS has dropped, but also a large portion of the density has transfered from high accelerated collapse regions to
free fall collapse regions. The non desorption mechanisms still present a multi-modal behavior, but with significantly smaller high density regions. 
Their distribution clearly highlights the non-linear way these mechanism act together or against each other. For $r$, 
the addition of informative prior information seems to reduce the HDS as well, centralizing the density, but still favoring slightly the 
production of H$_2$O against OH. Therefore, we conclude that the addition of gas phase species as a regularization factor outperforms the use 
of just a non-informative uniform prior distribution. The HDS is reduced at an average of $\sim12\%$, which indicates an equivalent constraint on the parameter
space. Section~\ref{sec-3.2} will discuss the statistical and numerical results of our analysis, while Section~\ref{sec-3.3} will discuss the astrophysical 
implications.  For both these Sections we shall only concentrate on the results of the informative prior case. 

\subsection{High Density Regions}
\label{sec-3.2}
The $n_H$ is clearly the most constrained physical parameter. The marginal density function reveals that most of the density is between 
$2.2$ and $5 \times 10^{4}\,cm^{-3} $. The $\zeta$ is constrained to values higher than $6 \times 10^{17} \, s^{-1} $, while the $G_{\circ}$ to values
lower than $4 \:Habing$. The HDR for the $fr$ stays between $20\%$ and $45\%$, while the $C_f$ has 1 distinct HDR between $0.5$ and $1.55$ and 
one long heavy tail between $2$ and $3$ times the default free fall rate. The $\epsilon$ presents two modes. The first HDR is between $0.4$ 
and $0.8$ and the second between $1.2$ and $1.4$. The marginal distribution of $\phi$, also presents two modes. One is centered 
around $10^5$. The second one is centered around $60$. The marginal distribution for $y$,  presents 2 disjoint high density regions as well. 
The first one indicates really low efficiency of about $10^{-6}$, while the second one a slightly higher $2 \times 10^{-3} - 8 \times 10^{-2}$. 
Finally, the distribution for the branching ratio parameter $r$ shows high density between $40\%$ and $70\%$ of Oxygen turning into ice water. 

In Figure~\ref{fig-3} we show the marginalized 2D PPD for our parameters. Note that the $n_H$ and the $fr$ are negatively dependent in 
a nearly linear way. On the other hand $G_{\circ}$ and $n_H$ seem to have a non-linear positive correlation, hitting a plateau after a 
certain gas density. Similarly, the $fr$ and the $C_f$ may have a clear peak, but also some evidence of a positive correlation. 
The relation between the cosmic ray desorption efficiency parameters, $\phi$ and $y$ reveals many distinct peaks throughout the domain space.
Note that the marginalized PPD for cosmic ray ionization rate and parameter $\phi$ shows a clear bimodal structure. However, focusing only on the denser
areas of the distribution we can observe a potential non linear correlation between the cosmic rays and the efficiency of the cosmic ray related parameter 
$\phi$. In general though, $\zeta$ is evidently a parameter that is not sufficiently constrained. This is already obvious by the 1D marginal distribution of $\zeta$, 
but the contrast of constrain between $\zeta$ and one of the most constraint parameters such as $n_H$ is depicted in Figure~\ref{fig-1}(6).

Due to the non-uniqueness of our solution space, examining the joint probability distribution of the PPD provides a useful insight. The dimensionality of the 
distribution makes a visualization impossible, so we chose to extract the statistical mean and the standard deviation for each one of the parameters from the 
most probable mode of the joint distribution. The joint distribution was approximated using a multivariate histogram and the most probable mode was chosen in 
a heuristic way and corresponds to $\sim 35\%$ HDR of the whole PPD. The values for the mean and standard deviation are given in Table~\ref{tbl-4}. 
As expected, the most probable mode of the joint PPD agrees with the HDR of the marginal parameter distributions. For the unimodal 1D marginalized 
distributions the most probable mode coincides completely, while for the multi-modal cases the most probable mode coincides with one of the modes. Hence,
purely based on the statistical interpretation we conclude that: a molecular cloud that matches the observed abundances should have low $n_H$,  
a low $fr$  and a low $G_{\circ}$. The $\zeta$ on the other hand is more likely to have high values, but the high standard deviation leaves room 
for significant variation. The collapse of the cloud may be insignificantly accelerated, while the branching ratio $r$ favors slightly the
branching into water, but with a high standard deviation. In terms of the non thermal desorption efficiency parameters,
we notice increased efficiency for all three of them. As a general result we conclude that the 9D space of the joint distribution has multiple peaks. 
Both the marginalized distributions and the denser peak of the joint distribution indicate that some of the parameters ($n_H$, $G_{\circ}, fr, C_f$) are well 
constrained, while other parameters($\zeta, r, \epsilon, \phi, y$) present possible variation that implies further astrophysical or statistical implications.

\subsection{Astrophysical Consequences}
\label{sec-3.3}

Here, we discuss our results for each of the parameters with regards to their astrophysical implication:

$\boldsymbol{n_H}$: The derived credible intervals for the gas density are in very good agreement with the properties of typical collapsing dark clouds, 
clumps and cores \citep{Mye1983,Ben1989,Bac2002,Ber2007}. Higher cloud densities ($>10^6\, cm^{-3}$), that are usually expected in hot cores after the cloud has 
collapsed \citep{Tak2004}, were explored, but showed nearly zero probability density in our analysis. 

$\boldsymbol{fr}$: Our study implies a depletion rate that is not high enough to dominate and is probably lower than $50\%$. \citet{Bac2002} suggest that 
freeze-out dominates when $n_H$ exceeds $\sim 3\times 10^4 \, cm^{-3}$ which is marginally the case in our study. When the freeze out dominates and densities
exceed $\sim 10^5 \, cm^{−3}$ , the abundance of CO ice is found to be significantly increased to typical gaseous values 
($\sim 10^{-4}$) \citep{Pon2006,Ber2007}. Furthermore, the ice water abundance is typically $5 \times 10^{-5}$ to $9 \times 10^{-5}$  and even higher 
at the highest densities \citep{Pon2005}. The ice CO and H$_2$O abundances in our case though, are about $0.5-1$ magnitude lower.
Therefore, along with the $n_H$ results, the lower freeze out rate estimated by our analysis can be explained by a different evolutionary stage of the
observed clouds. According to \citet{Fon2012} low depletion values can also imply a cloud that is going to form less massive objects.

$\boldsymbol{G_{\circ}}$: Our analysis showed that in order to match the observed ice abundances the $G_{\circ}$ is comparable to the 
standard interstellar radiation field of 1 $Draine$ or $\sim 1.7 \, Habing$ \citep{Dra1978}. 

$\boldsymbol{\zeta}$: In dense gas $\zeta$ is measured to be in the range of 1 to $5\times10^{-17} \, s^{-1}$ \citep{Ber1999}. However, considerable 
uncertainties have been reported in the literature
with derived values as high as $10^{-15} \, s^{-1}$, accounted to x-rays from a central source \citep{Dot2004}. These discrepancies may be due to whether $\zeta$
 is determined via  H$_3^+$ or HCO$^+$ measurements. Yet, \citet{Dal2006} claims that given latest evidence that the $\zeta$ range is 
 narrow and between $10^{-16}$ and $10^{-15}$, the question should be focused not on why the estimations are different, but on why they are so similar. 
Our analysis confirms $\zeta$ values higher than the typical estimations and is even consistent with the $10^{-16} \, s^{-1}$ estimations through the
H$_3^+$ determination. Most importantly, our study indicates high standard deviation on these values highlighting that such a variation should be 
expected. Theoretically, this is explained considering the fact that $\zeta$ lose energy while ionizing and exciting the gas through which they travel in conjunction 
with the possible variation in the origin of $\zeta$. Even though our astrochemical model does not account for the latter factors, our
probabilistic approach reflects their impact. 

$\boldsymbol{C_f}$: Our study shows that the collapse of the cloud should follow the expected free fall collapse. 
Higher $C_f$ values present moderate probability density, which implies that the observational constraints could potentially also be 
matched with different but also less likely sets of parameters (e.g. higher values for both $C_f$ and $fr$). 

$\boldsymbol{\epsilon, \phi, y}$: The desorption from H$_2$ efficiency parameter ($\epsilon$) estimates are significantly higher than the value 
reported by \citet{Rob2007} ($\epsilon<0.1$). The direct cosmic ray desorption efficiency ($\phi$), presents two peak values. One of them agrees with 
\citet{Rob2007} and is centered around $10^5$. The second one is centered around $60$, which is lower than the lowest limit studied by 
\citet{Rob2007}. For the cosmic ray-induced photodesorption efficiency ($y$), we have two probable estimates as well. The first one indicates 
really low efficiency. The second one presents a slightly higher efficiency that is still lower than the one estimated by \citet{Har1990} ($y=0.1$), 
but consistent with the results of \citet{Obe2009} for CO$_2$.  Our analysis in general indicates useful credible intervals for non thermal 
desorption efficiencies, highlighting though, that the reported non linearities can be tackled with further regularization factors such as 
molecule specific analysis and additional grain properties. Note as well that our astrochemical model does non include direct UV photodesorption 
which has recently be found to be efficient.

$\boldsymbol{r}$: The branching ratio proved to be a parameter with high but anticipated variability. Its marginal probability distribution presents 
the most statistically normal behavior with a mean that implies a shared branching ratio of Oxygen freezing into ice H$_2$O and 
ice OH, favoring slightly solid H$_2$O. The first laboratory experiment to reproduce the ice H$_2$O formation 
\citep{Dul2010} implied that the  hydrogenation of Oxygen is an important route for water formation. Furthermore, \citet{Caz2010} state 
that species such as OH are only transitory and quickly turn into ice water. However, they also state that  $\sim 30\%$ of the 
O coming on the grain is released in the gas phase as OH which can freeze back as H$_2$O. A pathway that  
is included in our model and  can explain both the high water abundance on the grains and the shared branching ratio $r$. At last the high 
branching ratio towards ice  OH highlights the importance of ice OH for the production of ice CO$_2$. 

\begin{figure}
\centering
\subfloat[]{
  \includegraphics[width=0.45\textwidth]{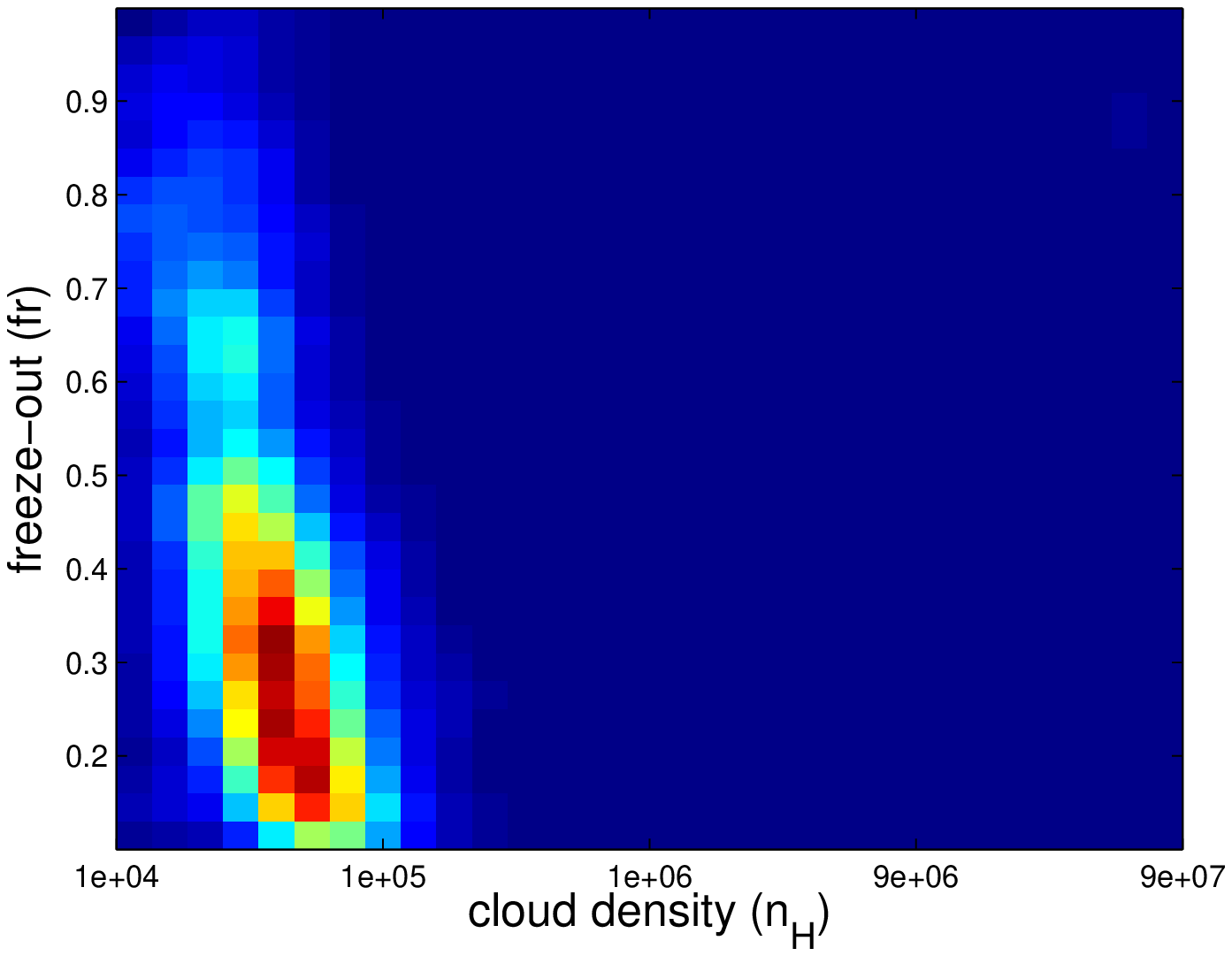}
}
\subfloat[]{
  \includegraphics[width=0.46\textwidth]{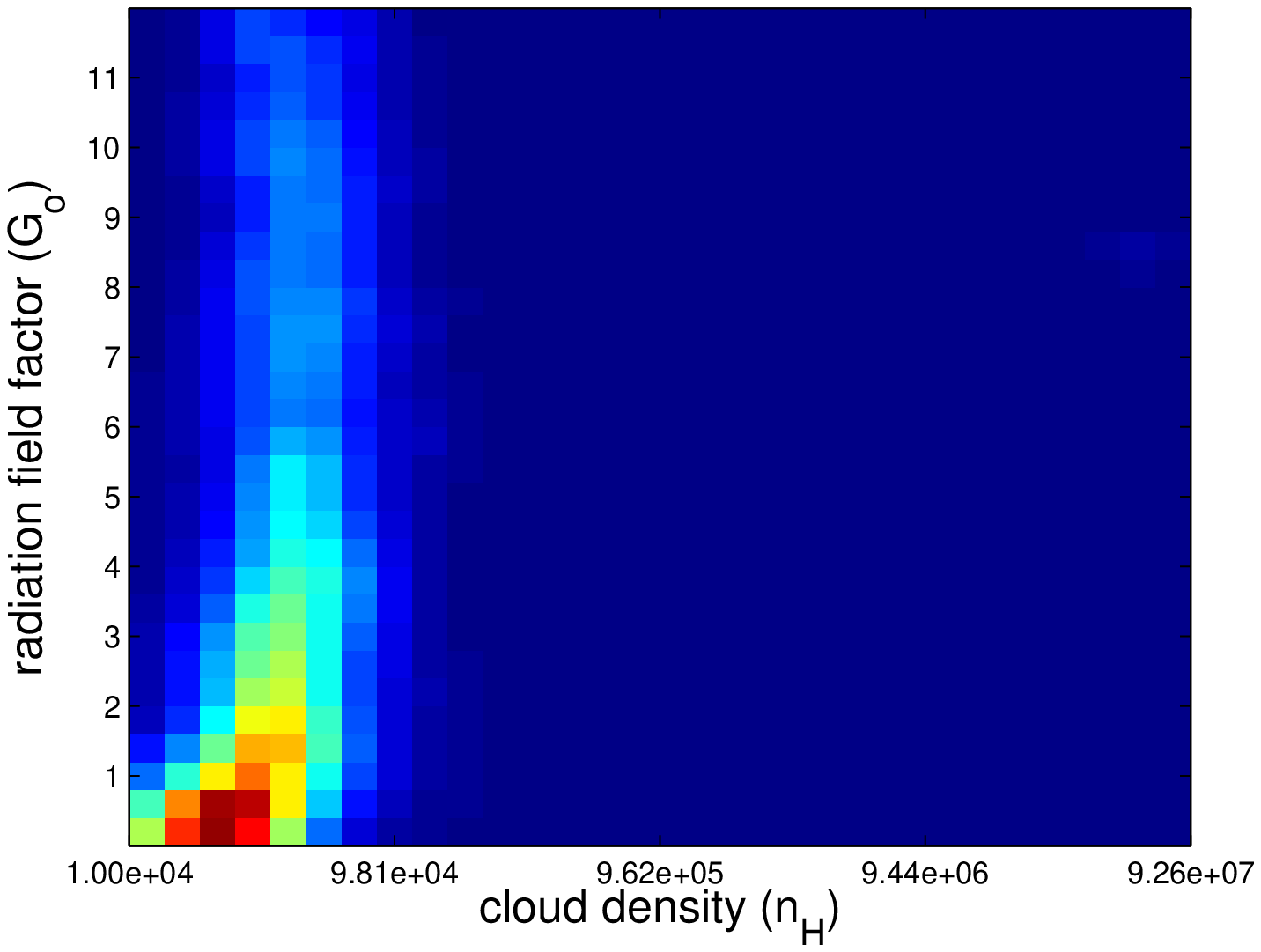}
}
\hspace{0mm}
\subfloat[]{
  \includegraphics[width=0.44\textwidth]{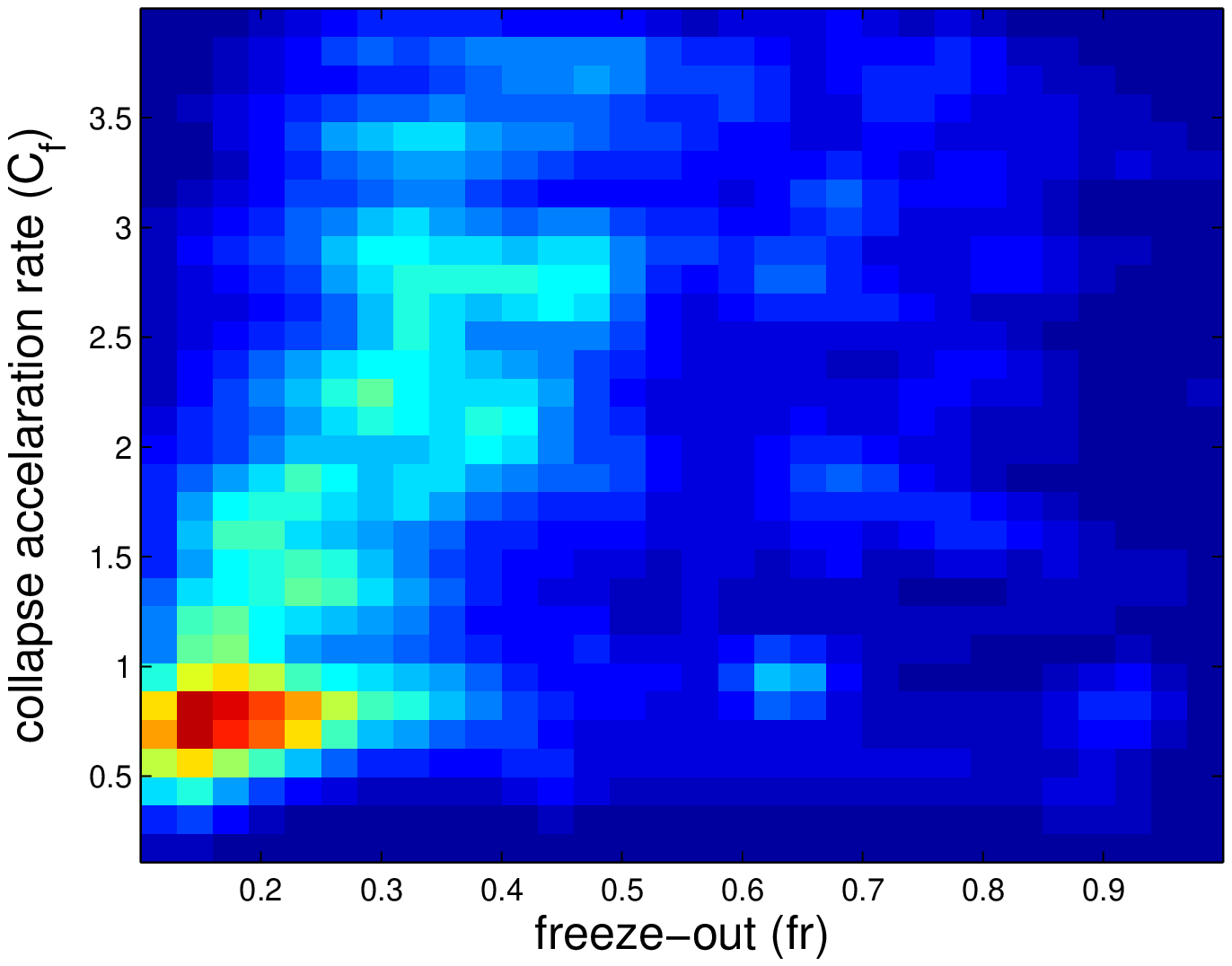}
}
\subfloat[]{
  \includegraphics[width=0.47\textwidth]{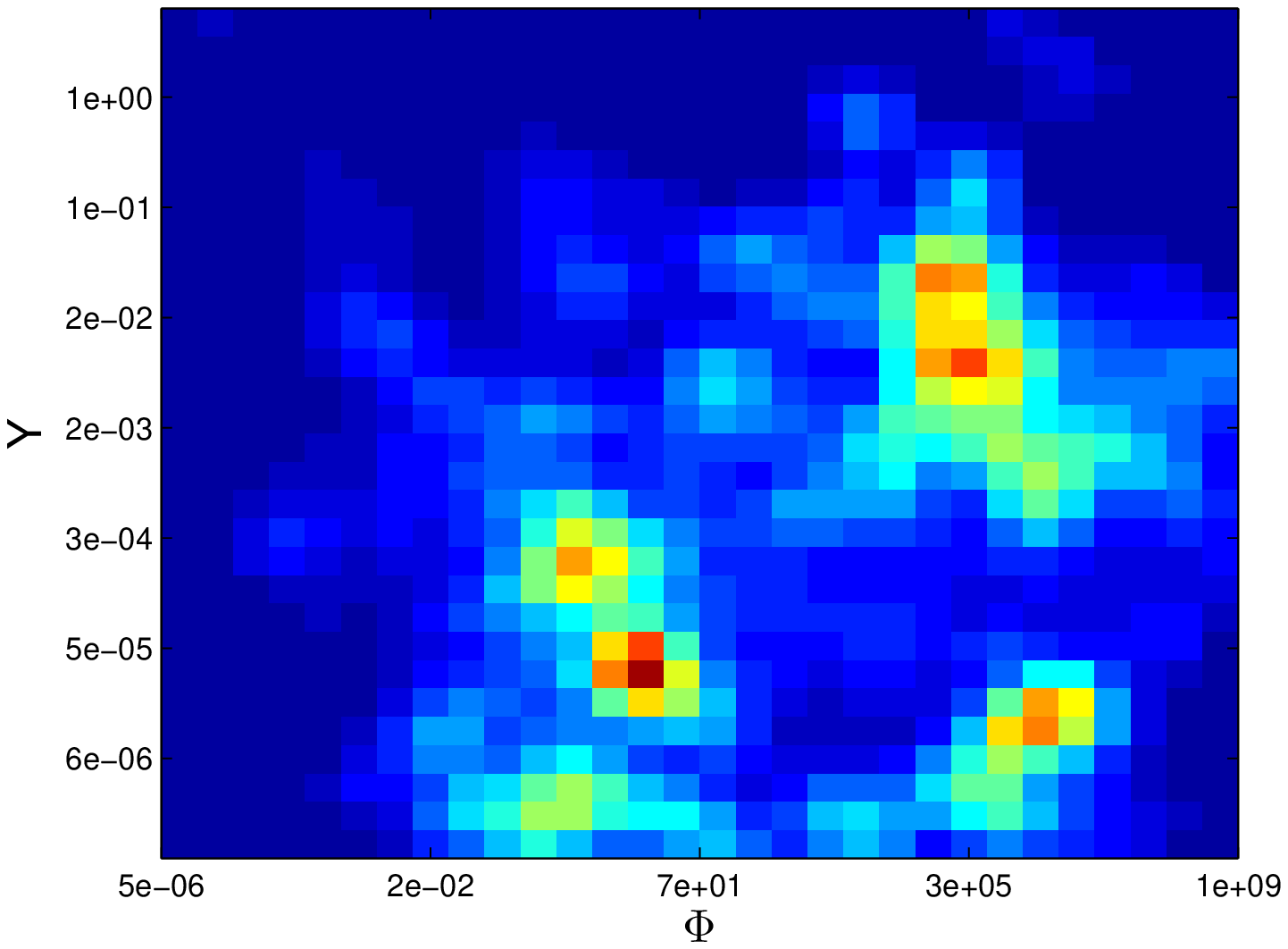}
}
\hspace{0mm}
\subfloat[]{
  \includegraphics[width=0.45\textwidth]{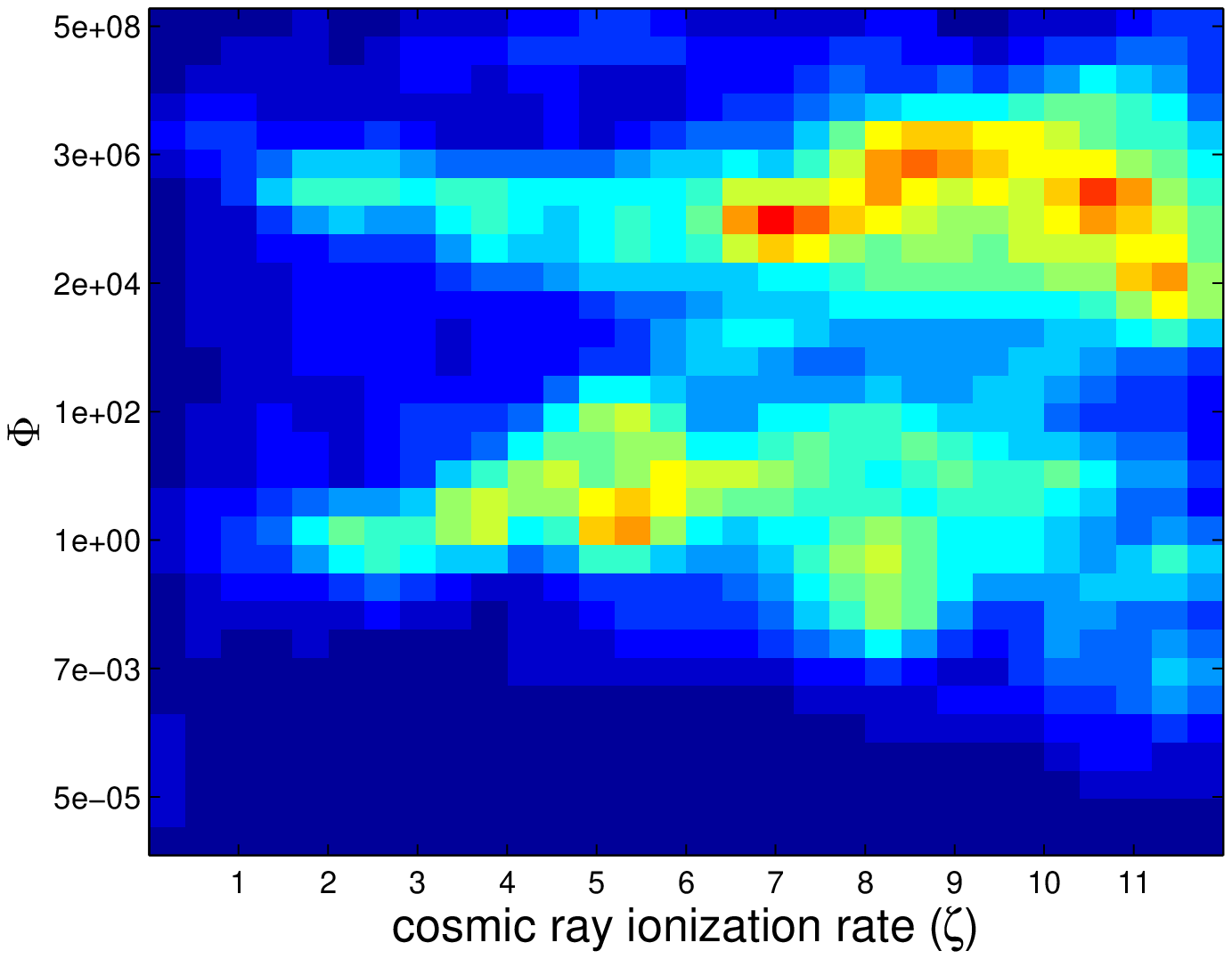}
}
\subfloat[]{
  \includegraphics[width=0.46\textwidth]{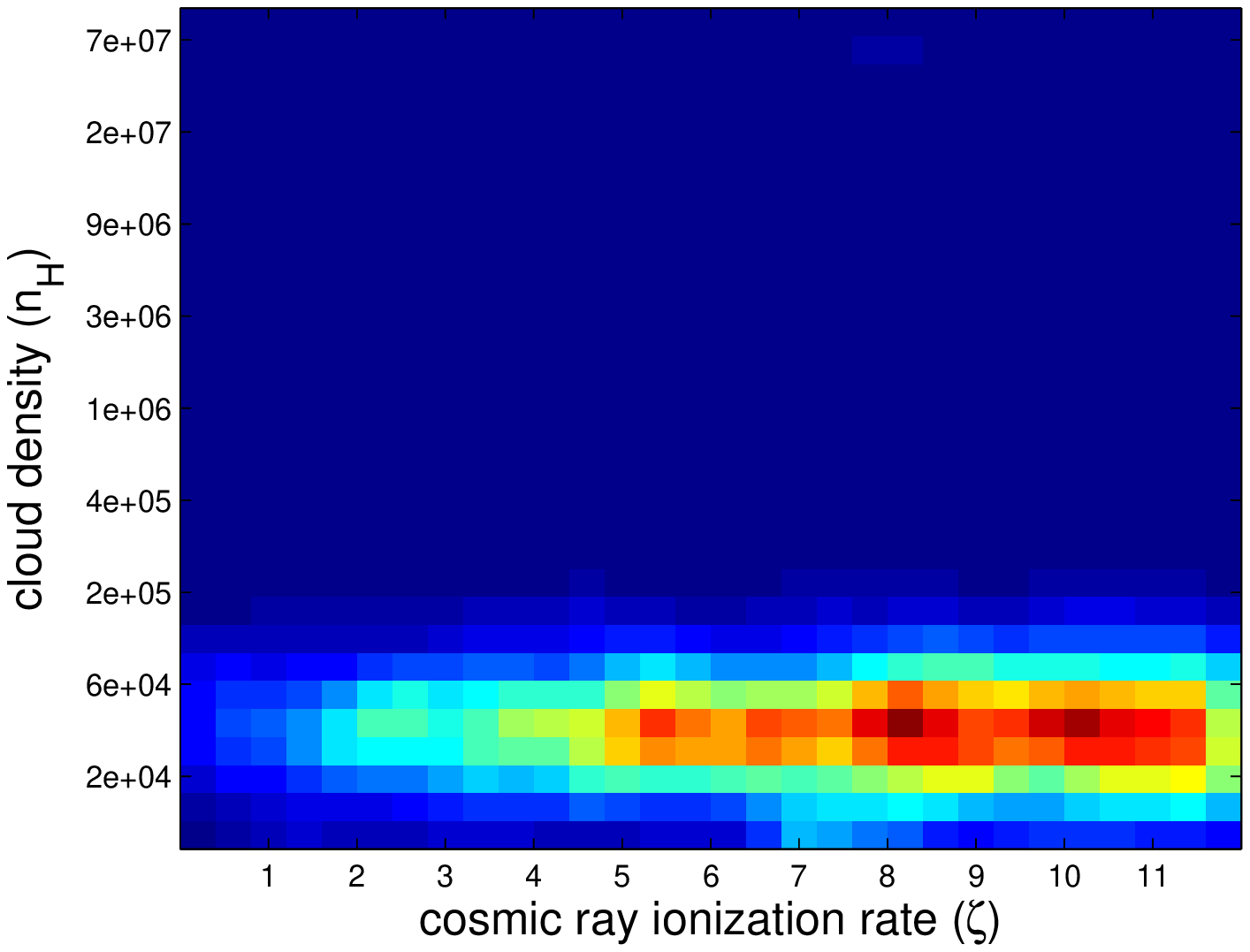}
}

\caption{2D marginalized posterior probability density functions.Warmer colors indicate higher probability density.\label{fig-3}}
\end{figure}

We now look at the correlation between our parameters as presented in Figure~\ref{fig-3}. The relation between the $n_H$ and depletion or freeze out 
has been the subject of many studies \citep{Bac2002, Chr2012, Fon2012, Hoc2014}. They all
conclude that the amount of depletion, the ice abundances and the density of the cloud should all scale together \citep{Raw1992}. 
Even though our analysis suggests a clear anti-correlation between $n_H$ and $fr$ (Figure ~\ref{fig-3}(a)), this result is completely in line with 
literature, since we are not analyzing the time evolution of the cloud, but instead focus on parameter fitting at specific time points. 
This negative correlation suggests that the less gas density we have the higher the depletion should be in order to match the observed ice abundances.
Our results are also in line with the negative 
correlation between depletion factor and $n_H$, derived by \citet{Fon2012} from CO observations. The positive correlation 
between $n_H$ and $G_{\circ}$ depicted in Figure~\ref{fig-3}(b) is confirming that the denser a cloud, 
the higher $\zeta$ values are needed to match the observations. The plateau after a density value ($\sim 7\times 10^4\, cm^{-3}$) indicates that the 
explored radiation field domain space is not high enough to penetrate the cloud after a density threshold. 
When $C_f$ is increased the freeze out timescale needs to be decreased since the final 
$n_H$ is reached quicker. This reduced timescale requires higher $fr$ values in order to simulate the observed ice abundances and this relation 
is depicted in Figure~\ref{fig-3}(c). Even thought not very straightforward, the relation between the cosmic ray desorption efficiency parameters, 
$\phi$ and $y$, is very interesting (Figure~\ref{fig-3}(d)). In most cases, the cosmic ray  photodesorption efficiency is either low or either high
for both direct cosmic ray heating and cosmic ray induced cases. However, there is a significant peak when the direct cosmic ray impact is very 
efficient,  whilst the cosmic ray induced impact is very inefficient.

\section{Conclusions}
\label{sec-4}
In this paper, we implemented a Bayesian MH parameter estimation analysis to solve a typical ill-posed inverse astrochemical problem. 
We have employed a chemical modeling code and solid phase observations in order to get a holistic insight into the behavior of physical and chemical 
parameters that drive ice chemistry in dark molecular clouds. The main conclusions of this work are as follows.
\begin{enumerate}
 \item The Bayesian method provides a systematic approach to solve nonlinear inverse problems with high noise levels and significant model uncertainties. 
 The MCMC technique allows to sample from complex probability distributions in an efficient way. As highlighted by our Blind Benchmark Test, we can conclude
  that the latter methods succesfully handle astrochemical ill-posed problems and reveal a more complete set of solution regions. On the contrary, single solution estimates 
 derived from traditional approaches would not have provided a complete picture of the solution space and would have contained a high risk of degeneracy.
 \item Our probabilistic approach to physical and chemical parameter estimation used a chemical network with deficiencies (especially for the grain part)
 and several assumptions. Nevertheless, the results both derived useful credible intervals
and highlighted model deficiencies implying even more promising results for tackling physical, chemical and model uncertainties for up to date models with targeted 
astrophysical goals.
 \item We confirm that the joint PPD of the solution space is highly non linear and multimodal and the 1D marginal PPD for each parameter are far from Gaussian 
 highlighting the complexity of the problem.
 \item Including abundances of gas phase species as a regularization factor and introduced as a Bayesian prior, increases the parameter constrain efficiency
 by $12\%$. This result can imply that observational regularization constraints compensate for any chemical code deficiencies. 
 Also, increasing the number of gas phase regularization factors will constrain even more the solution space. 
 \item We show that physical parameters such as $n_H$, $G_{\circ}$, $C_f$ are  highly constrained and their variation has a great impact on the derived 
 ice abundances.
 \item The high variation of $\zeta$ contradicts the theoretical $\zeta$ standard values in dense gas and indicates a larger credible interval instead.
 \item Non thermal desorption efficiencies act and counteract in a non linear way with each other or $\zeta$. This complex behavior should be analyzed 
 with extra regularization factors.
 \item Branching ratio parameters such as $r$ can be successfully estimated through Bayesian MCMC methods. Our results even though with high variability, 
 indicate that the detail or simplicity  of the dust grains chemical network can be encapsulated and reflected as certainty or uncertainty respectively.    
\end{enumerate}

\clearpage
\appendix

\section*{Appendix A\\Markov Chain Monte Carlo}
The MCMC framework uses a Markov chain to explore the parameter space and approximate the posterior probability distribution.  This chain consist of a series of 
states $\boldsymbol{\theta}^{(1)},...,\boldsymbol{\theta}^{(t)},...\boldsymbol{\theta}^{(T)}$, where the probability of $\boldsymbol{\theta}^{(t)}$ depends 
only on $\boldsymbol{\theta}^{(t-1)}$. MCMC methods require an algorithm for choosing states in the Markov chain in a random way. 
The MCMC sampler implemented for this paper was the Metropolis-Hastings (MH) algorithm. The MH is briefly outlined here using the following pseudocode:
\begin{enumerate}
\setlength{\itemsep}{0pt}
 \item Select a starting point $\boldsymbol{\theta}^{(1)}$ from the parameter space. Then for $i=2,3,... $ until convergence, repeat the following steps.
 \item Propose a random set of parameters according to a proposal distribution $q$, so that $\boldsymbol{\theta}^* \sim q(\boldsymbol{\theta}^i|\boldsymbol{\theta}^{i-1})$
 \item Calculate the posterior probability of the new parameters, $\pi(\boldsymbol{\theta}^*|\boldsymbol{\mathcal{Y}})$, using equation 2
 \item Accept the new parameters with probability 
 \begin{displaymath}
 \alpha(\theta^*|\theta^{i-1})=min\{1,\frac{q(\theta^{i-1}|\theta^*) \pi(\theta^*|\mathcal{Y})}{q(\theta^*|\theta^{i-1}) \pi(\theta^{i-1}|\mathcal{Y})}\}
 \end{displaymath}
 \item Calculate $u \sim Uniform (u;0,1)$
 \item if $u<a$ then accept the proposal, $\theta^i \leftarrow \theta^*$; otherwise, reject the proposal and $\theta^i \leftarrow \theta^{i-1}$
\end{enumerate}
The performance of the MH algorithm is highly dependent on the proposal distribution. The appropriate distribution should account for the complexity of the 
target distribution but it should still be computationally easy to draw samples from. In nonlinear problems such as ours, we expect a multimodal non Gaussian 
target joint distribution. Non gaussianity is not a problem for MCMC algorithms. However classical choices for the proposal distribution (i.e. Gaussian 
distribution) can potentially prevent the MCMC to converge to the target distribution, since the transition of the chain from one mode to another is not very 
possible. In our specific case, following former similar choices (eg. \citet{And1999,Baz2012}), and taking into account the characteristics of the expected target 
distribution, we chose the proposal distribution $q(\boldsymbol{\theta}^*|\boldsymbol{\theta}^t)$ to be a mixture of two Gaussians distribution centered at  
$\boldsymbol{\theta}^{(t)}$ and a uniform distribution on $\mathcal{D}_{{\theta}}$. Hence, for all parameters 
$\theta^*_k$ for k=1,..,9
\begin{align*}
\setlength{\itemsep}{0pt}
 \boldsymbol{\theta}^* &\sim  \mathcal{N}_{ \mathcal{D}_{\theta}}(\theta^t_k,\sigma^2_{k,1}) \text{with probability } 40\% \\
 \boldsymbol{\theta}^* &\sim  \mathcal{N}_{ \mathcal{D}_{\theta}}(\theta^t_k,\sigma^2_{k,2}) \text{with probability } 40\% \\
 \boldsymbol{\theta}^* &\sim  \mathcal{U}_{ \mathcal{D}_{\theta}} \text{with probability } 20\%
\end{align*}
The values for $\sigma^2_{k,1}$ and $\sigma^2_{k,2}$ were selected based on test runs. 
We run $m=8$ independent Markov chains of length $T=200000$. By using parallel and independent chains it is easier to understand 
the dependence of the MH performance on the initial parameter values guesses. Moreover, parallel chains provide insight on whether
convergence has been reached. Convergence was also decided based on empirical graphical aid. The length T of the chains was chosen 
confidently larger than the value of decided convergence.
In our case $q(\cdot)$ will be a symmetrical distribution. That means that $q(\boldsymbol{\theta}^{(t)}|\boldsymbol{\theta}^*) = q(\boldsymbol{\theta}^*|\boldsymbol{\theta}^{(t)})$ 
and the ratio in the acceptance probability $\alpha$ is simply the PPD ratio computed at $\boldsymbol{\theta}^*$ and $\boldsymbol{\theta}^t$. In simple words, that parameters that increase
the PPD are always accepted, while parameters that decrease the PPD are randomly accepted based on $\alpha$.

\pagebreak

\section*{APPENDIX B \\ UCL\_CHEM}
In recent years the molecular complexity of star forming regions has led in the development of complex, 
multi-point time dependent, gas-grain chemical and photon-dominated models which more accurately simulate the physics and 
the chemistry of the observed interstellar material. 
The chemical modeling code used in this paper is the UCL\_CHEM chemical code. A thorough description is given by  \citet{Vit2004}. 
UCL\_CHEM is a time and depth dependent gas-grain chemical model that can be used to estimate the fractional abundances 
(with respect to hydrogen) of gas and surface species in every environment where molecules are present. 
The model includes both gas and surface reactions and determines molecular abundances in environments where not only
the chemistry changes with time but also local variations in physical conditions lead to variations in chemistry.
Regardless of the object that is modeled, 
the code will always start from the most diffuse state where all the  gas is in atomic form and evolve the gas to its final 
density. Depending on the temperature, atoms and molecules from the gas freeze on to the grains and they 
hydrogenate where possible. The advantage of this approach is that the ice composition is not assumed but it is derived by a 
time-dependent computation of the chemical evolution of the gas-dust interaction process. 
The main categories for the physical and chemical input parameters are the initial elemental abundances, cosmic ray ionization rate ($\zeta$), 
radiation field strength ($G_{\circ}$), gas density ($n_H$), dust grain characteristics, freeze-out (species depletion rate), desorption processes 
and reaction database. 
The initial fractional elemental abundances, compared to the total number of hydrogen nuclei, were taken to be 
$0.14$, $4.0 \times 10^{-4}$ , $1.0 \times 10^{-4}$,
$7.0 \times 10^{-5}$, $1.3 \times 10^{-7}$, $1.0 \times 10^{-7}$ for helium, oxygen, carbon, nitrogen, sulphur and magnesium \citep{Sof2001}.
The gas phase network used by UCL\_CHEM is based on the UMIST database \citep{Mil2000}. Our chemical network also includes surface reactions as in \citet{Vit2004}. 
In total we have 208 species and 2391 gas and surface reactions included in our network
As an output, the code will compute the fractional abundances of all atomic and molecular species included in the network 
as a function of time.

\pagebreak

\section*{APPENDIX C\\ Exploring the Posterior Probability Density}
The MH simulations provide us with the joint parameter PPD. However, because of the high dimensionality of the distribution it is impossible 
to represent graphically the joint probability density. Therefore, we compute the marginal density for each parameter or for a subset of parameters by
integrating the PPD over the rest of the parameters except the ones we are interest in. For example, to obtain the joint marginal distribution of
$\boldsymbol{\theta}_a =\{d,fr\}$ we integrate over the rest of the parameters $\boldsymbol{\theta}_b = \{{\zeta,rad,bc, \epsilon, \phi,y,r}\}$,
\begin{displaymath}
\pi(\boldsymbol{\theta}_a|\boldsymbol{\mathcal{Y}}) = \int \pi(\boldsymbol{\theta}_a,\boldsymbol{\theta}_b|\boldsymbol{\mathcal{Y}})d\boldsymbol{\theta}_b
\end{displaymath}
The marginal probability distributions are visualized either with simple histograms for the case of univariate probabilities or 
with a bivariate histogram with intensity map for the case of bivariate probabilities.

Traditionally, in order to explore the posterior distribution, typical Bayesian estimates, such as the Posterior Mean are used.
However, for multi-modal and/or non Gaussian distributions the extraction of any useful estimator is most of the times meaningless. 
Instead, it is convenient to decrease the parameter space to High Density Regions (HDR) or credible intervals. HDR computation
and graphical representation is explained thoroughly by \citet{Hyn1996}. Following his paper we shortly define HDR as follows:

Let $f(x)$ be the density function of a random variable $X$. Then the $100(1-a)\%$ HDR is the subset $R(f_a)$ of the sample space of $X$ such that
\begin{displaymath}
 R(f_a) = {x:f(x) \geq f_a }
\end{displaymath}
where $f_a$ is the largest constant such that $Pr(X \in R(f_a)) \geq 1-a $

The above definition indicates two very important properties. From all the possible regions, HDR occupy the smallest possible volume and every point in the regions has 
probability density that is larger or equal than every point that doesn't belong in the regions.
HDR are very useful for analyzing and characterizing multi-modal distributions. In such cases, HDR might consist of several regions that are disjoint due to the number of modes. 
In the context of ice formation mechanisms these high density regions are very useful statistical outcomes of the Bayesian approach. Such regions 
provide us with a precise quantitative measure of how the ice and gas observations and their uncertainties impact the cloud parameters.

\acknowledgments

This work is supported by the IMPACT fund. The authors also acknowledge STFC for computational support and thank the anonymous referee for the useful suggestions 
that improved the paper.





\clearpage

\begin{deluxetable}{ccc}
\tablewidth{0pt}
\tablecaption{Parameter Definition Domain\label{tbl-1}}
\tablehead{
\colhead{Parameters $\boldsymbol{\theta}$} & \colhead{Unit}  & \colhead{Definition Domain $\mathbb{D}_{{\theta}}$}} 
\startdata
$\zeta$ & $10^{-17}\cdot s^{-1}$ &  1-10 \\ 
$G_{\circ}$  & $Habing$ &  1-10 \\ 
$n_H$ & $cm^{-3}$& $10^4 - 10^8$ \\ 
$fr$ &- &$0-100\%$ \\ 
$C_f$ & - & $0.5-3$ \\ 
$\epsilon$ & yield per H$_2$ formed & $0.01-1$ \\ 
$\phi$ &yield per cosmic ray impact &$10^2 - 10^6$ \\ 
$y$ & yield per photon &$10^{-3}-10^2$ \\ 
$r$ & - &$0-100\%$ \\ 
\enddata
\end{deluxetable}

\begin{deluxetable}{ccccccccc}
\tabletypesize{\footnotesize}
\tablecolumns{8} 
\tablewidth{0pc} 
\tablecaption{Observational Constraints (Average Fractional Abundances) \label{tbl-2}} 
\tablehead{
\multicolumn{4}{c}{Solid Phase Species} & \colhead{} & \multicolumn{3}{c}{Gas Phase Species} \\
\cline{1-4} \cline{6-8} \\
\colhead{H$_2$O} & \colhead{CH$_3$OH} & \colhead{CO} & \colhead{CO$_2$} & \colhead{} & \colhead{NH$_2$} & \colhead{N$_2$H$^+$} & \colhead{HCO$^+$}} 
\startdata
$7.47\pm 1.81 $ & $0.23\pm 0.13$ & $1.14\pm 0.84$ & $1.89\pm0.79$ & & $3.10\pm 2.24$ & $0.068\pm 0.049$ & $0.20\pm0.01$ \\ 
\enddata
\tablenotetext{*}{The fractional abundances are with respect to H nuclei}
\tablenotetext{**}{Solid phase abundances are in units of $10^{-5}$; Gas phase abundances are in units of $10^{-8}$ }
\end{deluxetable}

\begin{deluxetable}{ccc}
\tablewidth{0pt}
\tablecaption{Blind Benchmark Test\label{tbl-x1}}
\tablehead{
\colhead{Parameters $\boldsymbol{\theta}$} & \colhead{Unit}  & \colhead{Test Value}} 
\startdata
$\zeta$ & $10^{-17}\cdot s^{-1}$ &  2.4 \\ 
$G_{\circ}$  & $Habing$ &  2.6 \\ 
$n_H$ & $cm^{-3}$& $10^5$ \\ 
$fr$ &- &$42\%$ \\ 
$C_f$ & - & $1.3$ \\ 
$\epsilon$ & yield per H$_2$ formed & $0.02$ \\ 
$\phi$ &yield per cosmic ray impact &$150$ \\ 
$y$ & yield per photon &$0.1$ \\ 
$r$ & - &$75\%$ \\ 
\enddata
\end{deluxetable}

\begin{deluxetable}{ccc}
\tablecaption{High Density Spread.The lower the value of HDS the more constraint is a parameter.\label{tbl-3}}
\tablecolumns{3}
\tablewidth{0pt}
\tablehead{
\colhead{Parameters ${\theta}$} &\multicolumn{2}{c}{High Density Spread  HDS(\%)} \\
\colhead{} & \colhead{Non-Informative Prior} &
\colhead{Informative Prior} }
\startdata
$\zeta$ & $48$ &  $36$\\ 
$G_{\circ}$  & $46$ &  $30$ \\ 
$n_H$ & $16$& $09$ \\ 
$fr$ &$38$ &$28$ \\ 
$C_f$ & $45$ & $35$ \\ 
$\epsilon$ & $50$ & $42$ \\ 
$\phi$ &$33$ &$28$ \\ 
$y$ & $38$ &$33$ \\ 
$r$ & $43$ &$32$ \\ 
\enddata
\end{deluxetable}

\begin{deluxetable}{ccc}
\tablewidth{0pt}
\tablecaption{Mean and standard deviation for the most probable mode of the Joint Distribution. The mode corresponds to $\sim 35\%$ HDR of the 
total joint distribution.\label{tbl-4}}
\tablehead{
\colhead{Parameters $\boldsymbol{\theta}$} & \colhead{Unit}  & \colhead{Mean Value}} 
\startdata
$\zeta$ & $10^{-17}\cdot s^{-1}$ & $8.39 (\pm 2.8)$\\ 
$G_{\circ}$  & $Habing$ & $ 1.79 (\pm 1.27)$ \\ 
$n_H$ & $cm^{-3}$& $ 4.07 (\pm 2.34) \times 10^{4} $\\ 
$fr$ &- &$ 31 (\pm 21)\% $\\ 
$C_f$ & - & $1.18 (\pm 0.9) $\\ 
$\epsilon$ & yield per H$_2$ formed  & $0.52 (\pm 0.35) $\\ 
$\phi$ &yield per cosmic ray impact &$2.78(\pm 1.12) \times 10^6$ \\ 
$y$ & yield per photon &$ 3.35 (\pm 2.27) \times 10^{-3}$ \\ 
$r$ & - &$56 (\pm 23)\%$ \\ 

\enddata
\end{deluxetable}

\end{document}